\documentclass[10pt]{article}
\usepackage[dvips]{graphicx}
\usepackage[tight]{subfigure}
\usepackage[english]{babel}
\usepackage{amssymb}
\usepackage{amsmath}
\usepackage{blindtext}
\usepackage{geometry}
\def\be{\begin{equation}}
\def\ee{\end{equation}}
\def\bea{\begin{eqnarray}}
\def\eea{\end{eqnarray}}
\def\beaN{\begin{eqnarray*}}
\def\eeaN{\end{eqnarray*}}
\def\ed{\end{document}}
\def\bit{\begin{itemize}}
\def\eit{\end{itemize}}

\def\sig{\sigma}

\def\lam{\lambda}

\def\Del{\Delta}
\def\del{\delta}

\def\k{\kappa}
\def\alf{\alpha}

\def\di{\partial}

\def\Lix{\pounds_\xi}

\def\half{{\textstyle{1 \over 2}}}
\def\~{\tilde}
\def\lag{{{\cal L}}}
\def\m{\label}
\def\l{\left}
\def\r{\right}

\def\Bar{\overline}

\def\diag{\rm diag}

\usepackage{authblk}

\usepackage[dvipsnames]{xcolor}
\usepackage{graphicx}

\newcommand{\gog}{\mathfrak{g}}
\newcommand{\goh}{\mathfrak{h}}

\begin{document}

\title{ \bf  The field-theoretical formalism for TEGR}
\author[1,2]{E. D. Emtsova\thanks{Electronic address: \texttt{ed.emcova@physics.msu.ru}}}
\author[3]{A. N. Petrov \thanks{Electronic address: \texttt{alex.petrov55@gmail.com}}}

\affil[1]{Department of Physics, Ariel University, Ramat HaGolan Str. 65, Ariel, 40700, Israel}

\affil[2]{Department of Physics, Bar Ilan University, Max ve-Anna Webb Str. 1, Ramat Gan, 5290002, Israel}

\affil[3]{Sternberg Astronomical institute, MV Lomonosov Moscow State University
Universitetskii pr., 13, Moscow, 119234, Russia}

\date{\small \today}
\maketitle

\begin{abstract}
The teleparallel equivalent of general relativity (TEGR) is represented in a field-theoretical form, where tetrad and matter perturbations are propagated on a background solution of TEGR. Thus, the background tetrad and metric satisfy the background Einstein field equations. This presentation, where perturbations can be finite (not infinitesimal or approximate), is equivalent to the original form of TEGR.   The background can be arbitrary, usually corresponding to the solution of TEGR under consideration.  Such a formalism is Lagrangian based, where perturbations are classified as dynamic fields, and varying the Lagrangian with respect to dynamic variables leads to the equations for perturbations.  Gauge (inner) transformations are defined, and the gauge invariance of the Lagrangian and the field equations are stated. Applying the Noether theorem to the Lagrangian we construct conserved currents, related superpotentials and charges.  As an application, we have considered linear gravitational equations on the Ricci-flat background and analyzed their properties, finally deriving gravitational wave equations and the tetrad perturbations in TT-gauge.
By new formulae for charges, we have calculated the mass for the Schwarzschild and Kerr black holes and the angular momentum for the Kerr solution. The results are quite acceptable, which signals that the new formalism is powerful, can be used further, and has a potential for future development that can be applied to generalizations of TEGR.

\end{abstract}

\section{Introduction}
\m{Introduction}

The perturbative sector of teleparallel gravity has now developed into a broad and technically rich literature. Early work already addressed primordial perturbations in torsion-based formulations, showing that, in the teleparallel equivalent of general relativity (TEGR), inflationary observables can reproduce the standard general relativity (GR)  results once the vierbein is treated appropriately, while extensions to modified torsional theories immediately raise additional issues related to local Lorentz symmetry and the actual propagating content of the theory \cite{Wu:2011Teleparallelism,Izumi:2012Revisited,Wu:2016InflationTeleparallel}. In particular, the detailed cosmological-perturbation analysis of $f(T)$ gravity demonstrated that, at linear order around Friedmann-Lemaitre-Robinson-Walker (FLRW) backgrounds, the scalar and tensor sectors can look deceptively close to their GR counterparts, even though the nonlinear theory is known to carry extra structure \cite{Chen:2010fTPerturbations,Izumi:2012Revisited}.

A second line of development focused on cosmological perturbations in concrete torsion-based models relevant for late-time cosmology and structure formation. This includes the evolution of scalar perturbations in viable $f(T)$ models and their impact on the cosmological micro-wave background (CMB) and large-scale structure \cite{Nunes:2018StructureFormation}, as well as the full derivation of linear perturbation equations in both $f(T)$ and scalar-torsion $f(\phi)T$ models \cite{Golovnev:2018ModifiedTeleparallelPerturbations}. These studies made clear that teleparallel perturbation theory is highly sensitive to the choice of tetrad, to the treatment of the spin connection, and to the distinction between genuinely propagating modes and modes that only appear in incomplete or gauge-fixed formulations.

This issue motivated a series of works devoted to the covariant formulation and to the systematic treatment of the spin connection in perturbation theory. In particular, scalar cosmological perturbations in covariant $f(T)$ gravity were revisited by explicitly incorporating the spin connection and by separating diagonal and non-diagonal parts of the perturbed equations \cite{Toporensky:2020SpinConnection}. Closely related analyzes extended the discussion to scalar-torsion theories with non-minimal couplings and $f(\Phi,T)$-type dependence \cite{Toporensky:2022ScalarTorsionPerturbations}. In parallel, more general frameworks were developed for cosmological perturbations in teleparallel geometries themselves: one can formulate irreducible decompositions and gauge-invariant variables for the most general cosmologically symmetric teleparallel backgrounds \cite{Hohmann:2021GeneralCosmoPerturbations}, study the perturbations of $f(T,B)$ models where torsion and boundary-term contributions decouple in a characteristic way \cite{Bahamonde:2021BoundaryExtension}, and extend the analysis to non-flat FLRW backgrounds and to the most general homogeneous and isotropic tetrads \cite{Bahamonde:2023NonFlatFT}.

More recently, the perturbative analysis was pushed further toward genuinely broader torsion-based scalar-torsion frameworks. This includes the teleparallel analog of Horndeski gravity, for which the scalar-vector-tensor decomposition and the corresponding $\alpha$-parameter description were worked out up to second order \cite{Ahmedov:2023TeleparallelHorndeskiPerturbations}. At an even more general level, cosmological teleparallel perturbations were analyzed by perturbing not only the metric (or tetrad) but also the flat connection itself, with explicit applications to $f(T)$ models and with strong conclusions concerning strong-coupling and ghost issues on cosmological backgrounds \cite{Heisenberg:2024CosmologicalTeleparallelPerturbations}. Altogether, these works show that perturbations in torsion-based teleparallel gravity are not merely a reformulation of the metric case: the status of extra degrees of freedom, the role of the flat spin connection, and the dependence on the chosen background tetrad remain central structural questions.

Gravitational waves provide a parallel and equally active direction. Modified teleparallel theories such as $f(T)$, $f(T,B)$ and $f(T,T_G)$ were analyzed in detail from the viewpoint of wave propagation and polarization content \cite{Farrugia:2018GWModifiedTeleparallel,Abedi:2018GWModifiedTeleparallel}. These studies showed, for example, that pure $f(T)$ gravity does not generate extra polarizations at first perturbative order, whereas $f(T,B)$ theories can excite breathing and longitudinal scalar modes already at that level \cite{Farrugia:2018GWModifiedTeleparallel,Capozziello:2020WeakFieldFTB}. The multimessenger constraint from GW170817 was then incorporated directly into the $f(T)$ framework, where the speed of gravitational waves was shown to agree with the speed of light, while small deviations may still survive in the dispersion relation and damping sector \cite{Cai:2018GW170817fT}. The propagation of waves in the most general quadratic torsion-based second-order teleparallel theories was also investigated, revealing that the linearized theory effectively reduces to new general relativity and admits up to six possible polarizations, depending on the parameter choice \cite{Hohmann:2018GWPropagation}. Related weak-field analyses were further extended to higher-order teleparallel models and to the teleparallel analog of Horndeski gravity, where richer massive and massless polarization structures were identified \cite{Capozziello:2020HigherOrderTeleparallelGW,Bahamonde:2021TeleparallelHorndeskiGW}. More recently, the weak-field limit of new general relativity was revisited from the viewpoint of cosmological-perturbation theory in order to clarify the actual linearized degrees of freedom in the various branches of the theory \cite{Golovnev:2024GWNGR}.

The picture that emerges from this literature is therefore twofold. On the one hand, perturbations and waves in teleparallel gravity have been studied extensively in many concrete models, and several important technical lessons are now clear: the spin connection cannot be ignored in covariant formulations, the number of visible modes may depend strongly on the background and on the perturbative order, and cosmological and gravitational-wave sectors provide complementary probes of torsion-based modifications of gravity \cite{Golovnev:2018ModifiedTeleparallelPerturbations,Toporensky:2020SpinConnection,Heisenberg:2024CosmologicalTeleparallelPerturbations,Farrugia:2018GWModifiedTeleparallel,Hohmann:2018GWPropagation}.
However, most of the existing literature remains 1) model-specific, 2) depending on a concrete choice of the background  and 3) predominantly linearized from the start with the assumption that perturbations are considered infinitesimal. In addition, as a rule, 4) a self-consistent and full mechanism for constructing conserved quantities for perturbations was not suggested. 

In this respect, to close the above gaps one needs to develop systematically a field-theoretical formalism in the framework of teleparallel theories. Already, in more measure, it has been elaborated both for GR and other metric theories of gravity. The  metric perturbations and perturbations of other (matter) physical fields are propagated in a fixed background spacetime that may be curved, belong to  symmetries, or be flat. Such an approach of reforming a metric theory gives an exact and generally equivalent derivation with respect to the initial geometrical form of the theory. The main properties of field-theoretical formulations are
 \bit
 \item In the framework of the  field-theoretical formalism, unlike the usual geometrical derivation where the metric (spacetime) is a dynamic field, a spacetime is fixed. The configuration of dynamic fields (which are just the perturbations) propagates in this fixed background
spacetime. The background can be arbitrary but must satisfy the equations (background equations) of the initial theory. 

\item The field-theoretical method is Lagrangian based. By this, a) a so called dynamic Lagrangian (Lagrangian density)  is defined for the perturbations on a given background, which play a role of the dynamic fields;  b) field equations are defined by varying the action with respect to perturbations; c) conserved quantities are defined by applying the Noether theorem to the action with the dynamic Lagrangian. 

\item Equations and conserved quantities are covariant under coordinate transformations, preserving generally covariance properties of the initial form of geometrical theory. 

\item Gauge transformations (inner, non-coordinate) are defined explicitly with well described properties convenient in applications.

\item Because the field configuration is exact (without approximations and assumptions of infinitesimal quantities), gauge transformations and conserved quantities are exact as well. Following this, the possibility to construct approximations of all important expressions up to an arbitrary order is left and can be provided.
 \eit

Without hesitation, the pioneer work by Deser \cite{Deser_1970} can be considered as a start in the development of the field-theoretical method in GR and other metric theories. These ideas were developed significantly in the series of the  works by Deser and Tekin \cite{DT_2002,DT_2003,DT_2003a,DKT_2005,DT_2007,Tekin_2008} where  a construction
of conserved charges for perturbations about vacuum in various gravity theories in $D$ dimensions has been presented. The procedure for constructing such charges is close to that in \cite{AbbottDeser_1982} expanded to 
4D GR. Following \cite{AbbottDeser_1982}, the authors choose a symmetric vacuum, such as de Sitter (dS) or anti-de Sitter (AdS) spaces.

Another way of developing the work \cite{Deser_1970} has begun in the papers \cite{GPP_1984,PP_1988}, where Deser's approach is generalized to arbitrary curved backgrounds and arbitrary metric theories. Various applications and points of development in GR were presented \cite{Petrov_1992,Petrov_1993,Petrov_1995,Petrov_1997}. Next, the development of the field-theoretical approach in the metric modifications of GR (such as the Einstein-Gauss-Bonnet theory and others) with appropriate applications was also provided  \cite{Petrov_2005,Petrov_2009,Petrov_2011,KP_2013,PL_2013,Petrov_2018,Petrov_2019,Petrov_2021}. The main derivatives  of these results can be found in the book \cite{Petrov_KLT_2017} and the review \cite{PP_2019}.

It turns out that the field-theoretical method can be applied to arbitrary field theory, not necessarily to metric theory, see \cite{PP_1988,Petrov_KLT_2017}:
\bit
\item Perturbations (dynamic variables) are propagated on the background of fixed solutions of the theory.
\eit
In fact, teleparallel theories (for example, torsion based theories) relate to this class where the key variables are the tetrad components instead of the metric components in metric theories. To our knowledge, there are no attempts to represent teleparallel theories in the field-theoretical form including the simplest invariants of GR such as TEGR and symmetric TEGR. Therefore, the goal of the present work is to represent TEGR in a field-theoretical form generally without relations 1) to concrete models, 2) to choice of concrete backgrounds, 3) to assumptions of smallness of perturbations, and 4) with the constructions of conserved quantities. 

The paper is organized as follows. 

In Section \ref{FT.arbitrary}, the field-theoretical formalism is illustrated by an example of an arbitrary covariant field theory with the construction of field equations for perturbations, sources of linear expressions, and outlining properties of gauge invariance. 

In section \ref{FT_with_grav_theories},  the field-theoretical formalism is illustrated by an example of a general field theory, where the part of gravitational sector is accented especially. GR is  briefly considered in this context.

In section \ref{Elements.TEGR}, we outline the main properties of TEGR necessary in our presentation, including the derivation of field equations and comparison with GR. Besides, covariance with respect to coordinate transformations and local Lorentz rotations is considered.  

In section \ref{Dynamic_Lagrangian_TEGR},
we  construct the dynamic Lagrangian for the field-theoretical representation of TEGR with matter on the basis of torsion scalar. The dynamic variables are tetrad perturbations and matter perturbations.  Such a Lagrangian is given both in the exact form (without approximation) and in a quadratic approximation. Gauge (inner) transformations induced by both general covariance ({\em gauge transformations}) and local Lorentz rotations ({\em Lorentz gauge transformations}) are defined, and the invariance of the Lagrangian with respect to them is stated.

In section \ref{EofM_TEGR}, the equations of motion (field equations) for the field-theoretical representation of TEGR with matter are derived exactly (without approximations). Their invariance with respect to both gauge transformations and Lorentz gauge transformations is studied. A comparison of equations in the field-theoretical presentations of TEGR and GR is provided.

In section \ref{Conserved_quantities}, applying the Noether theorem to the dynamic Lagrangian we construct conserved currents, related superpotentials and charges for perturbations in the field-theoretical representation of TEGR with matter. All expressions are exact, without approximations. 

In section \ref{Applications}, we apply the theoretical results to describe linear equations and gravitational waves in the field-theoretical TEGR on the Ricci flat background. Gauge invariance and Lorentz gauge invariance is considered, and TT-gauge is derived in the terms of tetrad perturbations. Another application is a calculation of charges for the Schwarzschild and Kerr black holes by new formulae. The results turn out the expected ones. In the case of the Schwarzschild black hole, the acceptable mass was obtained starting from both the Schwarzschild coordinates and the isotropic coordinates, and both are compared with the classical results in GR. In the  Kerr black hole case,  the acceptable mass and angular momentum obtained starting from the Boyer-Lindquist coordinates, however, with two different choices of the tetrads. The results are compared with those calculated by another formalism without using a background structure. 

In section \ref{Concluding_remarks}, we summarize the obtained results and discuss ideas for possible further development of the formalism.

In Appendix \ref{Perturbation_ISC_TEGR}, we comment explicitly a possibility to introduce perturbation of the inertial
spin connection in the field-theoretical reformulation of TEGR; 
Appendices \ref{app:derivation_current} and \ref{app:superpotential_construction} help us better understand  the way conserved currents and related superpotentials are derived.

\section{The field-theoretical formulation for perturbations in an arbitrary field theory}
\label{FT.arbitrary}
\setcounter{equation}{0}

In the present section, we outline the field-theoretical approach in an arbitrary field generally covariant theory.

\subsection{Lagrangian and exact equations for perturbations} 

Consider an arbitrary field theory with a covariant Lagrangian of general form:
\begin{equation}\label{L_Q}
\lag = \lag(Q^A)\,,
\end{equation}
where a generalized variable $Q^A$ can contain variables of both the gravitational sector (it could be a metric, as an example) and the sector of matter sources (here, we consider arbitrary tensor densities). The
field equations for the system (\ref{L_Q}) are derived in a usual compact form as
\begin{equation}\label{EofM}
\frac{\delta \lag}{\delta Q^A} =0\,,
\end{equation}
where the Lagrange derivative is defined as
\begin{equation}\label{Lagrange.der}
\frac{\delta \lag}{\delta Q^A} \equiv \frac{\di \lag}{\di Q^A} - \di_\alf\l(\frac{\di \lag}{\di Q^A_{,\alf}} \r)+ \di_{\alf}\di_{\beta}\l(\frac{\di \lag}{\di Q^A_{,\alf\beta}}\r)\,,
\end{equation}
if Lagrangian contains derivatives up to a second order and is enough smooth.

Let us decompose $Q^A$ into the background part $\bar{Q}^A$
 and the dynamical part $q^A$,
perturbations,
\begin{equation}\label{decomposition}
Q^A=\bar{Q}^A+q^A.
\end{equation}
The background fields are fixed in the sense that they satisfy the background
equations
\begin{equation}\label{EofM_bar}
\frac{\delta \bar{\lag}}{\delta\bar{Q}^A} =0\,,
\end{equation}
where $ \bar{\lag}={\lag} (\bar{Q})$.

Following the recipe \cite{GPP_1984,PP_1988,Petrov_KLT_2017}, we construct a so-called {\em dynamical} Lagrangian for perturbations, as dynamical variables,
\begin{equation}\label{L_dyn}
   {\lag}^{dyn} (\bar{Q},q)= {\lag} (\bar{Q}+q)-q^A \frac {\delta \bar{\lag}}{\delta \bar{Q}^A}-\bar{\lag}\,.
\end{equation}
The background equations should not be taken into account before the variation of $ {\lag}^{dyn}$
 with respect to $\bar{Q}^A$. To obtain the field equations for the dynamical variables (perturbations) $q^A$ one
has to vary (\ref{L_dyn}) with respect to $q^A$,
\begin{equation}\label{EofM_dyn}
\frac{\delta {\lag}^{dyn}}{\delta q^A} =0.
\end{equation}
Let us show that this equations are equivalent to the original field equations (\ref{EofM}).
Using the evident property
\begin{equation}\label{property}
    \frac{\delta {\lag} (\bar{Q}+q)}{\delta \bar{Q}^A} =\frac{\delta {\lag} (\bar{Q}+q)}{\delta q^A}
\end{equation}
the field equation (\ref{EofM_dyn}) can be represented in the form:
\begin{equation}\label{EofM_and_bar}
\frac{\delta {\lag}^{dyn}}{\delta q^A} = \frac{\delta}{\delta \bar{Q}^A}[{\lag} (\bar{Q}+q)-\bar{\lag}]=0.
\end{equation}
This form shows that the equations for perturbations are equivalent to the equations
of the theory (\ref{EofM}) if the background equations (\ref{EofM_bar}) hold.

Defining the
“background source” as
\begin{equation}\label{source}
\theta^q_A\equiv \frac{\delta {\lag}^{dyn}}{\delta \bar{Q}^A} \equiv \frac{\delta {\lag}^{dyn}}{\delta q^A}-\frac{\delta}{\delta \bar{Q}^A} q^B \frac{\delta \bar{\lag}}{\delta \bar{Q}^B}\,,
\end{equation}
using the property (\ref{property}) and the equations (\ref{EofM_dyn}), one obtains another form for the
equations (\ref{EofM_dyn}):
\begin{equation}\label{EofM_dyn+}
     -\frac{\delta}{\delta \bar{Q}^A} q^B \frac{\delta \bar{\lag}}{\delta \bar{Q}^B} =\theta^q_A\,.
\end{equation}
Here and below, unless explicitly stated otherwise, the Lagrange derivative is always understood to act on the entire expression to its right.
The left hand side here is linear in perturbations, whereas the right hand side (the source of the linear side) is non-linear. Note that we do not use an assumption that perturbations have to be infinitesimal, they can be arbitrary and exact.

Nevertheless, the presentation (\ref{EofM_dyn+}) gives an algorithm of the consequent approximations of the field equations (equations of motion). Indeed, the Lagrangian (\ref{L_Q}) could be represented as a functional expansion \cite{DeWitt_1965,Petrov_KLT_2017}:
\begin{equation}\label{L_Q_exp}
     \lag= \bar \lag +q^A\frac{\delta \bar{\lag}}{\delta\bar Q^A}+\frac{1}{2!}q^A \frac{\delta }{\delta\bar Q^A} q^B \frac{\delta \bar{\lag}}{\delta\bar Q^B}+\frac{1}{3!}q^A \frac{\delta }{\delta\bar Q^A} q^B \frac{\delta}{\delta\bar Q^B} q^C \frac{\delta \bar{\lag}}{\delta\bar Q^C}+ \dots + {\rm div}
\end{equation}
One concludes that the dynamical Lagrangian is not less than quadratic in perturbations
\begin{equation}\label{dyn_exp}
     \lag^{dyn}= \frac{1}{2!}q^A \frac{\delta}{\delta\bar Q^A} q^B \frac{\delta \bar{\lag}}{\delta\bar Q^B}+\frac{1}{3!}q^A \frac{\delta }{\delta\bar Q^A} q^B \frac{\delta }{\delta\bar Q^B} q^C \frac{\delta \bar{\lag}}{\delta\bar Q^C}+ \dots + {\rm div}
\end{equation}
Then, the background source (\ref{source}) is not less than quadratic in perturbations as well, and the equations (\ref{EofM_dyn+}) can be rewritten in the form
\begin{equation}\label{EofM_dyn++}
     -\frac{\delta}{\delta \bar{Q}^A} q^B \frac{\delta \bar{\lag}}{\delta \bar{Q}^B} = \frac{1}{2!}\frac{\delta}{\delta \bar{Q}^A}    q^B \frac{\delta }{\delta\bar Q^B} q^C \frac{\delta \bar{\lag}}{\delta\bar Q^C}+\frac{1}{3!}\frac{\delta}{\delta \bar{Q}^A}q^B \frac{\delta}{\delta\bar Q^B} q^C \frac{\delta }{\delta\bar Q^C} q^D \frac{\delta \bar{\lag}}{\delta\bar Q^D} +\ldots
\end{equation}
The expressions (\ref{L_Q_exp})-(\ref{EofM_dyn++}), in principle, are exact (not approximate). In the case of smallness of perturbations and their derivatives (for example, $q^A \ll
 Q^A, etc.$) the expansions could be interrupted. Then the field equations, for example, can be rewritten in the approximate form
\begin{equation}\label{EofM_dyn+++}
     -\frac{\delta}{\delta \bar{Q}^A} q^B \frac{\delta \bar{\lag}}{\delta \bar{Q}^B} = \frac{1}{2!}\frac{\delta}{\delta \bar{Q}^A}    q^B \frac{\delta }{\delta\bar Q^B} q^C \frac{\delta \bar{\lag}}{\delta\bar Q^C}\,,
\end{equation}
where the source is quadratic in perturbations.

\subsection{Gauge invariance induced by general covariance}

By our assumption, the theory with Lagrangian (\ref{L_Q}) is covariant with respect to a diffeomorphism (that presents an external symmetries). Based on this property, we formulate a gauge invariance (inner symmetries) for only perturbations in the theory with the Lagrangian (\ref{L_dyn}). Then both the coordinates and the background are left unchanged. Now we recall the necessary properties of the Lie derivatives and the Lie displacements. First, 
\begin{equation}\label{exp_Lie}
       \exp\pounds_\xi= 1+\pounds_\xi+\frac{1}{2!}\pounds^2_\xi + \frac{1}{3!}\pounds^3_\xi+\ldots \,,
\end{equation}
where $\xi^\alf$ is enough smooth displacement vector.
If the quantity $Q^A$ (geometrical object) is a function of $P^A$ (another geometrical object, and their derivatives), that is, $Q^A=Q^A(P^B)$ then due to the Lie displacement formalism one has
\begin{equation}\label{Q=Q(q)}
       Q^A(\exp\pounds_\xi P^B)= \exp\pounds_\xi Q^A(P^B)\,.
\end{equation}
Then, if one substitutes $Q'^A = \exp\pounds_\xi Q^A $ into (\ref{L_Q}) one obtains
\begin{equation}\label{Q_Lprime}
       \lag(Q'^A )= \exp\pounds_\xi  \lag(Q^A )=\lag(Q^A ) +(\exp\pounds_\xi-1)\lag(Q^A ) = \lag(Q^A ) +{\rm div}\,.
\end{equation}
One obtains a divergence at the right hand side because $\lag$ is a scalar density of the weight +1. Then, variation of (\ref{Q_Lprime}) gives the same field equations as the variation of (\ref{L_Q}), thus reflecting the diffeomorphism invariance of the theory (\ref{L_Q}).  On the other hand, let us substitute $Q'^A = \exp\pounds_\xi Q^A $ into equations (\ref{EofM}), then we obtain
\begin{equation}\label{EofM_exp}
\frac{\delta \lag (Q'^A)}{\delta Q'^A} = \exp\pounds_\xi \frac{\delta \lag (Q^A)}{\delta Q^A}.
\end{equation}
Now, let us apply the decomposition like (\ref{decomposition}) to $Q'^A = \exp\pounds_\xi Q^A $:
\begin{equation}\label{Q+q}
(\bar Q'^A + q'^A) = \exp\pounds_\xi (\bar Q^A + q^A)\,.
\end{equation}
It is also a result of Lie displacement, where both  geometric objects $\bar Q^A$ and $q^A$ are transformed under the Lie displacement. It is the result of external transformations.

Now, let us reformulate the transformation (\ref{Q+q}) in the form where both coordinates and background quantities $\bar Q^A$ are not changed, but only the dynamic quantities $q^A$ are changed as
\begin{equation}\label{gauge_q}
q'^A = q^A + (\exp\pounds_\xi -1) (\bar Q^A + q^A)\,.
\end{equation}
It is internal transformations, not external ones. Therefore, we call them gauge transformations. 1) Let us substitute (\ref{gauge_q}) into (\ref{L_dyn}) and take into account the properties (\ref{Q=Q(q)}) - (\ref{Q+q}). Then
\begin{equation}\label{gauge_L}
\lag^{dyn}(\bar Q,q') = \lag^{dyn}(\bar Q,q) + (\exp\pounds_\xi -1) \lag(\bar Q + q) - \l((\exp\pounds_\xi -1)(\bar Q^A + q^A)\r)\frac{\delta\bar \lag}{\delta\bar Q^A}\,.
\end{equation}
One can see that the dynamical Lagrangian is gauge invariant up to a divergence (it is the second term in (\ref{gauge_L})) and if the background equations (\ref{EofM_bar}) hold. 2) Now, substituting  (\ref{gauge_q}) into (\ref{EofM_dyn}) and taking into account properties (\ref{Q=Q(q)}) - (\ref{Q+q}), one obtains
\begin{equation}\label{gauge_EofM}
\frac{\delta\lag^{dyn}(\bar Q,q')}{\delta q'^A} =\frac{\delta\lag^{dyn}(\bar Q,q)}{\delta q^A}+  (\exp\pounds_\xi -1)\l(\frac{\delta\lag^{dyn}(\bar Q,q)}{\delta q^A} + \frac{\delta\bar \lag}{\delta\bar Q^A}\r)\,.
\end{equation}
Thus, if the field equations (\ref{EofM_dyn}) hold, they are gauge invariant with respect transformations (\ref{gauge_q}) in the case if the background equations (\ref{EofM_bar}) hold as well. 3) At last, let us substitute (\ref{gauge_q}) into (\ref{source}) and obtain after not so prolonged transformations
\begin{equation}\label{gauge_source}
\theta^{q'}_A = \theta^{q}_A +  (\exp\pounds_\xi -1)\l(\frac{\delta\lag^{dyn}(\bar Q,q)}{\delta q^A} + \frac{\delta\bar \lag}{\delta\bar Q^A}\r) - \frac{\delta}{\delta \bar Q^A} \l[\frac{\delta\bar \lag}{\delta\bar Q^B}(\exp\pounds_\xi -1)(\bar Q^B + q^B) \r]\,,
\end{equation}
where in the last term the function $(\exp\pounds_\xi -1)(\bar Q^B + q^B)$ is not varied.
One can see that in spite of the field equations (\ref{EofM_dyn}) and the background equations (\ref{EofM_bar}) hold the source is not gauge invariant. In metric theories this fact reflects non-localization of the gravitational energy.

Finally, we must stress that the field-theoretical derivation does not assume that the values of perturbations $q^A$ are smaller than the values of background quantities $\bar Q^A$. In fact, they can be quite compatible and all the expressions of the field-theoretical re-presentation of the theory are finite, not infinitesimal. However, the smallness of $q^A$ can be assumed, and following expansions can follow.

\section{The field-theoretical formulation of gravitational theories}
\label{FT_with_grav_theories}
\setcounter{equation}{0}

Let us consider the Lagrangian (\ref{L_Q}) as the Lagrangian of an arbitrary  gravitational theory, where the gravitational part is separated from the part of matter sources:
\begin{equation}\label{L_metric}
{\lag(Q^A)}= {-\frac{1}{2 \kappa}} \lag^G(g^{\cal A})+\lag^M(g^{\cal A}, \Phi^{{\cal A}'}).
\end{equation}
Thus, $Q^A = \{g^{\cal A}, \Phi^{{\cal A}'}\}$, $\kappa = 8\pi $ in units $G=c=1$, Italic capital letters mean generalized indices for the gravitational variables, Italic capital letters with prime mean generalized indices for matter variables. Concerning gravitational variables, they could be $g^{\cal A} = \{ g_{\mu\nu},~g^{\mu\nu},~\sqrt{-g}g^{\mu\nu},~\dots \} $ in metric presentation, where Greek letters mean spacetime coordinate indices denoting $0,1,2,3$; $g \equiv {\rm det}g_{\mu\nu}$; in scalar-tensor theories $g^{\cal A}$ is presented by metric and scalar field, etc. In the tetrad presentation, it could be  $g^{\cal A} = \{e^a{}_\mu,~e_a{}^\mu, ~\ldots \}$, where small Latin indices denoting $0,1,2,3$ mean tetrad components for the tetrad vectors $e^a{}_\mu$, etc.; $e \equiv\sqrt{-g} \equiv {\rm det}\,e^a{}_\mu$. 

Following (\ref{decomposition}) we define perturbations as $h^{\cal A}$ and $\phi^{{\cal A}'}$ by the decompositions
\begin{eqnarray}
 g^{\cal A}& =& \bar g^{\cal A} +h^{\cal A},
\label{g_dec} \\
\Phi^{{\cal A}'} &=&  \bar\Phi^{{\cal A}'} + \phi^{{\cal A}'}\,. 
\label{Phi_dec}    
\end{eqnarray}
Then the system of equations (\ref{EofM_dyn+}) is separated onto the gravitational and matter
parts:
\begin{equation}\label{G_equations}
    {G}^L_{\cal A} + \Phi^L_{\cal A} = 2 \kappa \theta_{\cal A}^{tot}\,,
\end{equation}
\begin{equation}\label{M_equations}
     \Phi^L_{{\cal A}'} = \theta_{{\cal A}'}^M\,.
\end{equation}
In this equations linear in perturbations left hand sides are defined as follows. The left hand side of (\ref{G_equations}) is divided into two parts, the pure gravitational part: 
\begin{equation}\label{E_a}
       {G}^L_{\cal A} \equiv \frac{\delta}{ \delta \bar{g}^{\cal A}} {h}^{\cal B} \frac{\delta\bar{\lag}^G}{\delta \bar{g}^{\cal B}}\,,
\end{equation}
and a part based on the background matter Lagrangian $\bar\lag^M$:
\begin{equation}\label{Phi_a}
     \Phi^L_{\cal A}  \equiv - 2 \kappa \frac{\delta}{ \delta \bar{g}^{\cal A}}\Big(\goh^{{\cal B}} \frac{\delta\bar{\lag}^M}{\delta \bar{\gog}^{{\cal B}}}+\phi^{{\cal A}'} \frac{\delta\bar{\lag}^M}{\delta \bar{\Phi}^{{\cal A}'}}\Big)\,.
\end{equation}
The left hand side in (\ref{M_equations}) is defined as
\begin{equation}\label{Phi_A}
     \Phi^L_{{\cal A}'}  \equiv - \frac{\delta}{ \delta \bar{\Phi}^{{\cal A}'}} \phi^{{\cal B}'} \frac{\delta\bar{\lag}^M}{\delta \bar{\Phi}^{{\cal B}'}}\,.
\end{equation}
The sources in (\ref{G_equations}) and (\ref{M_equations}) are defined as follows
\begin{equation}\label{EMT}
       \theta_{{\cal A}}^{tot} \equiv \frac{{\delta{\lag}^{dyn}(\bar g,\bar \Phi, h,\phi)}}{\delta\bar g^{{\cal A}}}\,,
\end{equation}
\begin{equation}\label{EMT+}
       \theta_{{\cal A}'}^{M} \equiv \frac{{\delta{\lag}^M_{dyn}(\bar g,\bar \Phi, h,\phi)}}{\delta\bar \Phi^{{\cal A}'}}\,,
\end{equation}
where ${\lag}^{dyn}(\bar g,\bar \Phi, h,\phi)$ is the dynamic Lagrangian  totally related to (\ref{L_metric}) and
where ${\lag}^M_{dyn}(\bar g,\bar \Phi, h,\phi)$ is the dynamic Lagrangian related to the matter part in (\ref{L_metric}). However, we do not concretize ${\lag}^M(g,\Phi)$ in (\ref{L_metric}) and below are concentrated more on the gravitational sector.

We assume, anyway, that the gravitational sector directly is coupled to  a spacetime, which is considered as an arena for evolution of physical interactions. Thus, we can connect the variables of the gravitational sector $g^{{\cal A}}$ and $\bar g^{{\cal A}}$, for example, with the contra-variant spacetime metric $g^{\mu\nu}$ and $\bar g^{\mu\nu}$, respectively.  For simplicity, we assume that these variables are connected by non-singular algebraic relations, for example $g^{{\cal A}} = g^{{\cal A}}(g^{\mu\nu})$, $\bar g^{{\cal A}} = \bar g^{{\cal A}}(\bar g^{\mu\nu})$. Thus,
to obtain the gravitational equations defined in a background spacetime, one has to contract (\ref{G_equations})
with ${\partial  \bar{g}^{{\cal A}}}/{ \partial \bar{g}^{\mu\nu}} $:
\begin{equation}\label{G_eqs_munu}
      {G}^L_{\mu\nu} + \Phi^L_{\mu\nu} =  \kappa t_{\mu\nu}^{tot}\,.
\end{equation}\
Here, the left hand side is linear in ${h}^{{\cal A}}$ and $\phi^{{{\cal A}}'}$, the right hand side is the total symmetric (metric) energy-momentum tensor density for perturbations:
\begin{equation}\label{EMT++}
       t_{\mu\nu}^{tot} \equiv 2\frac{{\delta{\lag}^{dyn}(\bar g,\bar \Phi, h,\phi)}}{\delta\bar g^{\mu\nu}} \equiv t_{\mu\nu}^{g} +t_{\mu\nu}^{m} \,,
\end{equation}
where the pure gravitational energy-momentum $t_{\mu\nu}^{g}$ is defined by $\lag^G$ and its variable only.
 Equations (\ref{G_eqs_munu}) can be obtained directly by variation  with respect to $\bar g^{\mu\nu}$ of the dynamical Lagrangian related to (\ref{L_metric}) with decomposition (\ref{g_dec}) and (\ref{Phi_dec}).

It is instructive to consider general relativity (GR) to illustrate the above.  Then, we choose  the Hilbert Lagrangian as a pure gravitational Lagrangian in (\ref{L_metric}):
\begin{equation}\label{L_GR}
\lag(g,\Phi) = {-\frac{1}{2 \kappa}} \sqrt{-g}R+\lag^M(g, \Phi),\,
\end{equation}
where $R$ is scalar curvature constructed with the use of $g_{\mu\nu}$.
It is a fact that among the metric gravitational variables, authors in many cases prefer to use a metric density $\gog^{\mu\nu} = \sqrt{-g}g^{\mu\nu}$ as a basic variable.  Thus, Fock \cite{Fock_1964} uses it to significantly simplify the GR equations applying the de Donder (harmonic) conditions; Deser \cite{Deser_1970} uses a perturbation of $\gog^{\mu\nu}$  to obtain finite (closed) expansions of the Einstein equations. Thus, in (\ref{L_GR}),  we set $g_{\alf\beta} = g_{\alf\beta}(\gog^{\mu\nu})$ and use the decomposition
\begin{equation}\label{gog_dec}
\gog^{\mu\nu} = \bar\gog^{\mu\nu}+ \goh^{\mu\nu}\,
\end{equation}
for constricting the field-theoretical formulation with  gravitational perturbations $ \goh^{\mu\nu}$.

Then, the related to (\ref{L_GR}) dynamical Lagrangian has a form
 \be
 {\lag}^{\rm dyn}  =
 {\lag}(\bar\gog+\goh,\,\bar \Phi+\phi) -
 \goh^{\mu\nu}\frac{\del \bar{\lag}}{\del \bar{\gog}^{\mu\nu}}
-
\phi^{{\cal A}'} \frac{\delta \bar {\lag}}{\delta \bar\Phi^{{\cal A}'}}
 -
 \bar{\lag}\,.
 \m{b-b11}
 \ee
 Its variation with respect to $\bar g^{\mu\nu}$ leads just to equations of the form (\ref{G_eqs_munu}). For simplicity we restrict ourselves  by the Ricci-flat background $\bar R = 0$, where  dynamic matter is also absent. Then the equations take the form
\begin{equation}\label{G_eqs_munu_g}
      {G}^L_{\mu\nu} ( \goh) =  \kappa t_{\mu\nu}^{g}( \goh)\,,
\end{equation}\
  where $G^L_{\mu\nu}$ is expressed as
 \be  G^L_{\mu\nu}( \goh) \equiv
\frac{\delta}{\delta\bar{g}^{\mu\nu}} \goh^{\rho\sig}
\frac{\delta{(\sqrt{-\bar g} \bar R)}}{\delta \bar{\gog}^{\rho\sig}}
\equiv  \half \l(\bar\nabla_\rho\bar\nabla^\rho
\goh_{\mu\nu} + {\bar g_{\mu\nu}}\bar\nabla_{\rho}\bar\nabla_{\sig}
\goh^{\rho\sig} - \bar\nabla_{\rho}\bar\nabla_{\nu}\goh_{\mu}^{~\rho} - \bar\nabla_{\rho}\bar\nabla_{\mu}\goh_{\nu}^{~\rho}\r)\, ,
 \m {b-b21}
 \ee
  $\bar\nabla_\mu$ is the covariant derivative constructed with $\bar g_{\mu\nu}$. Also, we derive a pure gravitational energy-momentum tensor density \cite{GPP_1984,Petrov_KLT_2017}:
 \be
{t}^g_{\mu\nu}( \goh) \equiv \frac{2\delta \lag^{dyn}}{\delta \bar g^{\mu\nu}}\Big|_{\lag^M=0} = {1 \over \k} \l[\sqrt{-\bar g}
\l(-\del^\rho_\mu \del^\sig_\nu +\half \bar g_{\mu\nu} \Bar
g^{\rho\sig}\r)\l(\Del^\alf{}_{\rho\sig}\Del^\beta{}_{\alf\beta} -
\Del^\alf{}_{\rho\beta}\Del^\beta{}_{\alf\sig}\r) + \bar\nabla_\tau
Q^\tau{}_{\mu\nu}\r]\, ;\m{b-b24}
 \ee
 \bea 2 Q^\tau{}_{\mu\nu}( \goh) &\equiv & -\bar g_{\mu\nu}
\goh^{\alf\beta}\Del^\tau{}_{\alf\beta}+  \goh_{\mu\nu}
\Del^\tau{}_{\alf\beta}\bar g^{\alf\beta}- \goh^\tau{}_{\mu}
\Del^\alf{}_{\nu\alf}- \goh^\tau{}_{\nu} \Del^\alf{}_{\mu\alf}\nonumber \\ && +
\goh^{\beta\tau}\l( \Del^\alf{}_{\mu\beta}\bar g_{\alf\nu} +
 \Del^\alf{}_{\nu\beta}\bar g_{\alf\mu}\r)
\nonumber \\ && +   \goh^{\beta}{}_{\mu}\l( \Del^\tau{}_{\nu\beta}-
 \Del^\alf{}_{\beta\rho}\bar g^{\rho\tau}\bar g_{\alf\nu}\r) \nonumber \\ &&
+\goh^{\beta}{}_{\nu}\l( \Del^\tau{}_{\mu\beta}-
 \Del^\alf{}_{\beta\rho}\bar g^{\rho\tau}\bar g_{\alf\mu}\r)\, .
\m{b-b25}
 \eea
Here, the tensor $\Delta^\rho{}_{\mu\nu}$ is a perturbation of the Christoffel symbols
 \be
\Del^\alpha{}_{\mu\nu} \equiv \Gamma^\alpha{}_{\mu\nu} - \Bar
{\Gamma}^\alpha{}_{\mu\nu} = \half g^{\alf\rho}\l( \bar\nabla_\mu g_{\rho\nu}
+ \bar\nabla_\nu g_{\rho\mu} - \bar\nabla_\rho g_{\mu\nu}\r)\, , 
\m{b-b13}
 \ee
and depends on $\goh^{\mu\nu}$ through the decomposition (\ref{gog_dec}): $g_{\mu\nu} = g_{\mu\nu}(\gog)= g_{\mu\nu}(\bar\gog + \goh)$. 
One can see that ${t}^g_{\mu\nu}$ is not less than quadratic in  gravitational variables under expansions.

\section{Elements of TEGR}
\label{Elements.TEGR}
\setcounter{equation}{0}

The Lagrangian in TEGR has a form \cite{Aldrovandi_Pereira_2013}
\be
\lag =   \frac{e}{{2}\kappa}T  + \lag^M(g^{\cal A},\Phi^{{\cal A}'})\,,
\m{lag+M}
\ee
where the Lagrangian of the gravitational sector
is defined by the torsion scalar
\be
T
\equiv
\frac{1}{4} {T}{}^\rho{}_{\mu\nu} {T}_\rho{}^{\mu\nu} + \frac{1}{2} {T}{}^\rho{}_{\mu\nu} {T}{}^{\nu\mu}{}_\rho - {T}{}^\rho{}_{\mu\rho} {T}{}^{\nu\mu}{}_\nu\,
\m{T.scalar}
\ee
with ${T}{}^\rho{}_{\mu\nu} = e_a{}^\rho{T}{}^a{}_{\mu\nu}$. The torsion tensor ${T}{}^a{}_{\mu\nu}$ is defined as
\begin{equation}\label{tor}
{T}{}^a{}_{\mu\nu} = \di_\mu e^a{}_\nu - \di_\nu e^a{}_\mu + {\omega}{}^a{}_{c\mu}e^c{}_\nu - {\omega}{}^a{}_{c\nu}e^c{}_\mu
\end{equation}
with the tetrad components $e^a{}_\nu$
 connected to the metric by
\begin{equation}
    g_{\mu\nu}=\eta_{ab} e^a{}_\mu e^b{}_\nu\,,
\m{g_munu}
\end{equation}
where the Latin indices are the tetrad ones, $\eta_{ab} = \diag (-1,1,1,1)$
and $e= \det e^a{}_\nu$.
It is useful to recall that the transformation of tetrad indices into spacetime ones and vice versa is performed by contraction with tetrad vectors, for example,  ${T}{}^\rho{}_{\mu\nu} = e_a{}^\rho{T}{}^a{}_{\mu\nu}$ here, etc. For simplicity, we assume that $\lag^M$ depends on $e^a{}_\nu$ only through $ g_{\mu\nu}$.

In TEGR and its modifications,  the teleparallel affine connection (a so called Weitzenb\"ock connection) ${\stackrel{\scriptscriptstyle W}{\Gamma}}{}^\alpha {}_{\kappa \lambda}$ plays a crucial role. It is flat, i.e., the related curvature is equal to zero:
\begin{equation}\label{RiemTEGR}
    {\stackrel{\scriptscriptstyle W}{R}}{}{}^\alpha {}_{\beta \mu \nu} \equiv \di_\mu {\stackrel{\scriptscriptstyle W}{\Gamma}}{}^\alpha {}_{\beta\nu} - \di_\nu {\stackrel{\scriptscriptstyle W}{\Gamma}}{}^\alpha{}_{\beta \mu} + {\stackrel{\scriptscriptstyle W}{\Gamma}}{}^\alpha{}_{\kappa \mu}{\stackrel{\scriptscriptstyle W}{\Gamma}}{}^\kappa {}_{\beta \nu} - {\stackrel{\scriptscriptstyle W}{\Gamma}}{}^\alpha{}_{\kappa \nu}{\stackrel{\scriptscriptstyle W}{\Gamma}}{}^\kappa {}_{\beta \mu}=0\,.
\end{equation}
The connection ${\stackrel{\scriptscriptstyle W}{\Gamma}}{}^\alpha {}_{\kappa \lambda}$ is also compatible with the physical metric, that is, the corresponding non-metricity is zero: ${\stackrel{\scriptscriptstyle W}{Q}}{}   {}_{\mu \alf \beta}  \equiv  {\stackrel{\scriptscriptstyle W}{\nabla}}{}_\mu g_{\alf \beta}=0$, where the covariant derivative ${\stackrel{\scriptscriptstyle W}{\nabla}}{}_\mu$ corresponds to ${\stackrel{\scriptscriptstyle W}{\Gamma}}{}^\alpha {}_{\kappa \lambda}$\,.

The torsion in definition (\ref{tor}) is covariant with respect to both coordinate transformations and local Lorentz rotations. Indeed, first, it can be rewritten in the evidently coordinate covariant form:
\begin{equation}\label{tor.cov}
{T}{}^a{}_{\mu\nu} = \nabla_\mu e^a{}_\nu - \nabla_\nu e^a{}_\mu + {\omega}{}^a{}_{c\mu}e^c{}_\nu - {\omega}{}^a{}_{c\nu}e^c{}_\mu\,,
\end{equation}
where $\nabla_\mu$ is a covariant derivative constructed using the Levi-Chivita connection $\Gamma^\alf{}_{\beta\gamma}$ with the metric $g_{\mu\nu}$.
Second,
the torsion (\ref{tor.cov})  is Lorentz covariant  due to the incorporation of
the inertial spin connection ${\omega}{}^a{}_{c\nu}$. It is defined as
\begin{equation}\label{ISCdef}
    \omega{}^a{}_{b\mu} = -e_b{}^\nu \stackrel{\scriptscriptstyle W}{\nabla}_\mu e^a{}_\nu.
\end{equation}
 Working in the covariant formulation of teleparallel gravity, the local Lorentz covariance means that the tensorial quantities are transformed covariantly under the simultaneous transformation of both the tetrad and the inertial spin connection:
\begin{equation}\label{e-rot}
e'^a {}_{\mu} = \Lambda^a {}_b (x) e^b {}_\mu\,,
\end{equation}
\begin{equation}\label{spin_trans}
\omega'{}^a {}_{b \mu}=\Lambda {}^a {}_c  (x) \omega{}^c {}_{d \mu} \Lambda {}_b {}^d   (x)  + \Lambda {}^a {}_c  (x) \partial_\mu \Lambda {}_b {}^c  (x) ,
\end{equation}
where $\Lambda {}^a {}_c  (x)$ is the matrix of a local Lorentz rotation, and $\Lambda {}_a {}^c  (x)$  is an inverse matrix  of the latter. The operation on the right hand side of (\ref{spin_trans}) tells us that the inertial spin connection can be equalized to zero by an appropriate local Lorentz transformation. Then, by another local Lorentz rotation it can be represented in the form:
\begin{equation}\label{telcon}
 {\omega}{}^a{}_{c\nu}=\Lambda^a {}_b \partial_\nu    \Lambda_c{}^b.
\end{equation}

Let us discuss the role of the inertial spin connection incorporated into the torsion tensor (\ref{tor}) or (\ref{tor.cov}). We rewrite the Lagrangian (\ref{lag+M}) in the other form \cite{Aldrovandi_Pereira_2013}:
\be
{\lag} =   \frac{e}{2\kappa}T +\lag^M= -\frac{e}{2\kappa}R +\lag^M -\frac{1}{\kappa}\di_\mu\l(e{T}{}^{\nu\mu}{}_\nu\r) \,,
\m{lag+div}
\ee
where $R$ is the Riemannian curvature scalar presented as a function of tetrad components in correspondence with (\ref{g_munu}), see \cite{Landau_Lifshitz_1975}. Of course, $eR$ can be reformulated in the metric form. Thus,  the TEGR Lagrangian contains the inertial spin connection in the divergence only. Then, first, varying the action with the Lagrangian (\ref{lag+M}), the same (\ref{lag+div}), with respect to the tetrad components one obtains the Euler-Lagrange equation
 \be
\frac{\delta \lag}{\delta e^a{}_\rho} \equiv \frac{\di g^{\mu\nu} }{\di e^a{}_\rho}\frac{\delta \lag}{\delta g^{\mu\nu}} = \frac{e}{\kappa} \l(G_a{}^\rho - \kappa T_a^{M\rho} \r)=0.
 \m{EM+}
 \ee
 Here the relation (\ref{g_munu}) and the standard definitions of the Einstein tensor and the matter energy-momentum were used:
 \bea
 G_{\mu\nu} &\equiv &\frac{1}{\sqrt{-g}}\frac{\delta (\sqrt{-g}R)}{\delta g^{\mu\nu}}\,,
 \m{E.tensor}\\
  T^M_{\mu\nu} &\equiv &\frac{2}{\sqrt{-g}}\frac{\delta \lag^M}{\delta g^{\mu\nu}}\,.
 \m{M.tensor}
  \eea
 Because the formalism developed here uses Lagrangian derivatives, which suppress divergences as in (\ref{lag+div}), the inertial spin connections do not contribute to the field equations and, thus, TEGR and GR are equivalent on the level of equations. Therefore, constructing the field equations in the field-theoretical presentation, it is quite permissible to exclude the inertial spin connections from consideration without destruction of the Lorentz covariance.

\section{Dynamic Lagrangian in TEGR}
\label{Dynamic_Lagrangian_TEGR}
\setcounter{equation}{0}

\subsection{The total (without approximations) dynamic Lagrangian }
\label{Dynamic_Lagrangian_TEGR+}

We now construct the field-theoretical representation of TEGR with matter.
As a starting point, we use the Lagrangian (\ref{lag+M}) with the torsion scalar (\ref{T.scalar}) built from the torsion (\ref{tor}) (equivalently, (\ref{tor.cov})).
As discussed above, the inertial spin connection enters the action only through a total divergence; see (\ref{lag+div}); therefore, in what follows, we keep ${\omega}{}^{a}{}_{b\mu}$ fixed and do not display it explicitly. In more detail a possibility of considering perturbation  of inertial spin connection in the field-theoretical formulation of TEGR is discussed in Appendix \ref{Perturbation_ISC_TEGR}.

We choose the generalized set of variables as
$Q^{A}=\{g^{\cal A},\Phi^{{\cal A}'}\}=\{e^{a}{}_{\mu},\Phi^{{\cal A}'}\}$,
and introduce perturbations by the decomposition (\ref{decomposition}),
\begin{equation}
e^{a}{}_{\mu}=\bar e^{a}{}_{\mu}+\varkappa^{a}{}_{\mu},
\qquad
\Phi^{{\cal A}'}=\bar\Phi^{{\cal A}'}+\phi^{{\cal A}'}.
\m{decomposition.TEGR}
\end{equation}
Then the dynamical Lagrangian is defined exactly as in the general construction (\ref{L_dyn}),
\begin{equation}
\lag^{dyn}(\bar e,\varkappa;\bar\Phi,\phi)
\equiv
\lag(\bar e+\varkappa,\bar\Phi+\phi)
-\varkappa^{a}{}_{\mu}\,\frac{\delta\bar\lag}{\delta\bar e^{a}{}_{\mu}}
-\phi^{{\cal A}'}\,\frac{\delta\bar\lag}{\delta\bar\Phi^{{\cal A}'}}
-\bar\lag,
\m{dyn.lag.kappa.phi}
\end{equation}
where $\bar\lag=\lag(\bar e,\bar\Phi)$.
Using (\ref{lag+M}) 
the above definition can be written in advanced form:
\begin{equation}
\begin{aligned}
\lag^{dyn}
&=
\frac{1}{2\kappa}\Big[eT(\bar e+\varkappa)
-\varkappa^{a}{}_{\mu}\,\frac{\delta}{\delta\bar e^{a}{}_{\mu}}
\Big({\bar e}\,\bar T\Big)
-{\bar e}\,\bar T\Big]
\\
&\quad+
\Big[\lag^{M}(\bar e+\varkappa,\bar\Phi+\phi)
-\varkappa^{a}{}_{\mu}\,\frac{\delta\bar\lag^{M}}{\delta\bar e^{a}{}_{\mu}}
-\phi^{{\cal A}'}\,\frac{\delta\bar\lag^{M}}{\delta\bar\Phi^{{\cal A}'}}
-\bar\lag^{M}\Big],
\end{aligned}
\m{dyn.lag.kappa.phi+}
\end{equation}
where $\bar T=T(\bar e)$ and $\bar\lag^{M}=\lag^{M}(\bar e,\bar\Phi)$.
Subtracted zero-order in perturbations terms in (\ref{dyn.lag.kappa.phi}) and (\ref{dyn.lag.kappa.phi+}) are defined by the background Lagrangian 
\be
{\bar\lag} =   \frac{\bar e}{2\k}\bar T +\bar\lag^M= -\frac{\bar e}{2\k}\bar R +\bar\lag^M +{\rm div} \,.
\m{lag+div.bar}
\ee
Subtracted linear in perturbations terms in (\ref{dyn.lag.kappa.phi}) and (\ref{dyn.lag.kappa.phi+}) are  proportional to the background equation operators. An arbitrary curved background is defined by the background equations 
\begin{equation}\label{EofM_bar_T}
\frac{\delta \bar{\lag}}{\delta \bar e^a{}_\mu} =\frac{1}{\kappa}\bar e\,(\bar G_{a}{}^{\mu}-\kappa\,\bar T^{M}{}_{a}{}^{\mu}) =0,\qquad \frac{\delta \bar{\lag}^M}{\delta\bar{\Phi}^{{\cal A}'}} =0\,.
\end{equation}
that follows from (\ref{lag+div.bar}) and
which are not taken into account before variation of the dynamical Lagrangian, see (\ref{dyn.lag.kappa.phi}) and (\ref{dyn.lag.kappa.phi+}).

Gauge transformations for perturbations are inherited from the general rule (\ref{gauge_q}). 
The background fields are fixed in the sense that they satisfy the background
equations (\ref{EofM_bar_T}). 
For the displacement vector $\xi^{\alpha}$ the internal (gauge) transformations induced by general covariance are the specialization of (\ref{gauge_q}) to tetrad and matter variables read
\begin{equation}
\varkappa'{}^{a}{}_{\mu}
=
\varkappa^{a}{}_{\mu}
+\big(\exp\pounds_{\xi}-1\big)\big(\bar e^{a}{}_{\mu}+\varkappa^{a}{}_{\mu}\big),
\qquad
\phi'{}^{{\cal A}'}
=
\phi^{{\cal A}'}
+\big(\exp\pounds_{\xi}-1\big)\big(\bar\Phi^{{\cal A}'}+\phi^{{\cal A}'}\big).
\m{gauge.kappa.phi}
\end{equation}
and the corresponding transformation of $\lag^{dyn}$ defined in \eqref{dyn.lag.kappa.phi+} is the particular case of (\ref{gauge_L}). Substituting the above transformations into $\lag^{dyn}$ and using
(\ref{Q=Q(q)})--(\ref{Q_Lprime}) exactly as in Sec.~\ref{FT.arbitrary}, we obtain the TEGR version of (\ref{gauge_L}):
\bea
\lag^{dyn}(\bar e,\bar\Phi;\varkappa',\phi')
&=&
\lag^{dyn}(\bar e,\bar\Phi;\varkappa,\phi)
+\big(\exp\Lix-1\big)\,\lag(\bar e+\varkappa,\bar\Phi+\phi)
\nonumber\\
&-&\Big[\big(\exp\Lix-1\big)\big(\bar e^{a}{}_{\mu}+\varkappa^{a}{}_{\mu}\big)\Big]
\frac{\delta\bar\lag}{\delta\bar e^{a}{}_{\mu}}
-\Big[\big(\exp\Lix-1\big)\big(\bar\Phi^{{\cal A}'}+\phi^{{\cal A}'}\big)\Big]
\frac{\delta\bar\lag}{\delta\bar\Phi^{{\cal A}'}} .
\label{gauge.LT}
\eea
Since $\lag$ is a scalar density, the second term in (\ref{gauge.LT}) is a total divergence
(cf. (\ref{Q_Lprime})); therefore, if the background equations (\ref{EofM_bar_T}) hold,
then $\lag^{dyn}$ is gauge invariant up to a divergence:
\be
\lag^{dyn}(\bar e,\bar\Phi;\varkappa',\phi')
=
\lag^{dyn}(\bar e,\bar\Phi;\varkappa,\phi)+{\rm div}.
\m{dyn'=dyn}
\ee

The Lagrangian TEGR (\ref{lag+div}), being generally covariant, is also invariant with respect to Lorentz rotations (\ref{e-rot}). We repeat that because inertial spin connections are included in the divergence in (\ref{lag+div}) only, we do not consider them at all. Then, together with the gauge transformations (\ref{gauge.kappa.phi}), we introduce other gauge (inner) transformations 
\begin{equation}
\varkappa^*{}^{a}{}_{\mu}
=
\varkappa^{a}{}_{\mu}
+\big(\Lambda^a{}_b(x) -\delta^a_b \big)\big(\bar e^{b}{}_{\mu}+\varkappa^{b}{}_{\mu}\big),
\qquad
\phi^*{}^{{\cal A}'} =
\phi^{{\cal A}'}\,
\m{gauge.Lorentz}
\end{equation}
based on the local Lorentz invariance,
where $\hat\Lambda=\Lambda^a{}_b=\Lambda^a{}_b(x)$ is the matrix of local Lorentz rotations.
We choose only that the fields $\phi^{{\cal A}'}$ are tensor densities in a background spacetime and the Lorentz rotations do not act on them. The main property used to check the invariance of $\lag^{dyn}$ under transformations (\ref{gauge.Lorentz}) is
\begin{equation}
\lag(\bar e +\varkappa^*,\bar\Phi+\phi^*) = \lag(\hat\Lambda(\bar e +\varkappa),\bar\Phi+\phi) = \lag(\bar e +\varkappa,\bar\Phi+\phi)+\text{div}\,.
\m{lag.tot.Lorentz}
\end{equation}
Then, the dynamic Lagrangian (\ref{dyn.lag.kappa.phi}) is transformed under transformations  (\ref{gauge.Lorentz}) as
\begin{equation}
\lag^{dyn}(\bar e,\varkappa^*;\bar\Phi,\phi^*) = \lag^{dyn}(\bar e,\varkappa;\bar\Phi,\phi)
-\big(\Lambda^a{}_b -\delta^a_b \big)\big(\bar e^{b}{}_{\mu}+\varkappa^{b}{}_{\mu}\big)\,\frac{\delta\bar\lag}{\delta\bar e^{a}{}_{\mu}}+\text{div}\,
.
\m{dyn.lag.Lorentz}
\end{equation}
Thus, $\lag^{dyn}$ is gauge invariant up to a divergence under  transformations  (\ref{gauge.Lorentz}) if the background equations (\ref{EofM_bar_T}) hold.

To distinguish the gauge invariance induced by the general covariance and the gauge invariance induced by the Lorentz invariance, we call the first of them simply as {\em gauge invariance}, whereas the second one as {\em Lorentz gauge invariance}. Finally, one can check that the gauge transformations (\ref{gauge.kappa.phi}) and the Lorentz gauge transformations (\ref{gauge.Lorentz}) commute as applied to $\lag^{dyn}$ if the background equations (\ref{EofM_bar_T}) hold as well.

\subsection{Quadratic Lagrangian in TEGR}

In many applications, linear equations of the theory are considered. Therefore, a consideration of a quadratic approximation of the Lagrangian in the framework of the field-theoretical formulation of TEGR becomes very important. Besides, namely, the quadratic Lagrangian leads to constructing the energy-momentum of linear perturbations. Thus, returning to (\ref{dyn_exp}),  we set $Q^A = \{e^a_{\mu},\Phi^{{\cal A}'} \}$, and, suppressing ``div'', derive
\begin{equation}\label{dyn_exp2}
     \lag^{dyn}_2= \frac{1}{2!}q^A \frac{\delta }{\delta\bar Q^A} q^B \frac{\delta \bar{\lag}}{\delta\bar Q^B}  = \frac{1}{2!}\varkappa^a{}_\mu \frac{\delta }{\delta\bar e^a{}_\mu} \varkappa^b{}_\nu  \frac{\delta \bar{\lag}}{\delta\bar e^b{}_\nu}+\varkappa^a{}_\mu \frac{\delta }{\delta\bar e^a{}_\mu} \phi^{{\cal A}'}  \frac{\delta \bar{\lag}^M}{\delta\bar \Phi^{{\cal A}'}}+ \frac{1}{2!} \phi^{{\cal A}'}  \frac{\delta }{\delta\bar \Phi^{{\cal A}'}} \phi^{{\cal B}'}  \frac{\delta \bar{\lag}^M}{\delta\bar \Phi^{{\cal B}'}}  \,.
\end{equation}
Here, it is more interesting and important the first term related to gravitational variables $\varkappa^a{}_\mu$ only, and we analyze it in detail. Therefore 
we do not concretize the matter sources defined by $\lag^M$, and set $\phi^{{\cal A}'}=0$. Thus, do not derive two last terms in (\ref{dyn_exp2}). However, for the sake of generality, we can continue to consider a propagation of gravitational perturbations on an arbitrary curved background. Thus, the first term in (\ref{dyn_exp2}) is rewritten as
\begin{eqnarray}\label{dyn.2.kappa+}
     \lag^{dyn}_2(\varkappa)&=& \frac{1}{2!}\varkappa^a{}_\mu \frac{\delta }{\delta\bar e^a{}_\mu} \varkappa^b{}_\nu\frac{\delta }{\delta\bar e^b{}_\nu}  \Big( -\frac{\bar e}{2\k}\bar R+\bar\lag^M \Big)=\frac{1}{2!}\varkappa^a{}_\mu \frac{\delta }{\delta\bar e^a{}_\mu} {\cal K}^{\alf\beta}\frac{\delta }{\delta\bar\gog^{\alf\beta}}  \Big( -\frac{\bar e}{2\k}\bar R+\bar\lag^M \Big) \nonumber\\
     &=&
     -\frac{1}{4\k\bar e}\l( {\cal K}^{\rho\sig} - \frac{1}{2}\bar g^{\rho\sig}{\cal K}^\pi{}_\pi\r)\frac{\delta}{\delta\bar g^{\rho\sig}}
     {\cal K}^{\alf\beta}\Big[\bar R_{\alf\beta} - \k\Big(\bar T^M_{\alf\beta}- \frac{1}{2} \bar g_{\alf\beta}\bar T^M\Big)\Big]\,,
\end{eqnarray}
where the redefinition of $e^a{}_{\mu}$ is presented by
\be
\mathcal{K}^{\alpha\beta}(\varkappa,\bar e)
\equiv \varkappa^{b}{}_{\nu}\frac{\di\gog^{\alpha\beta} }{\di\bar e^b{}_\nu} =
\bar e\,
\varkappa^{b}{}_{\nu}\Big(\bar e_{b}{}^{\nu}\,\bar g^{\alpha\beta}
-\bar e_{b}{}^{\alpha}\,\bar g^{\nu\beta}
-\bar e_{b}{}^{\beta}\,\bar g^{\nu\alpha}
\Big)\,.
\label{K_def}
\ee
One can see that $\mathcal{K}^{\alpha\beta}$ is a tensor (invariant with respect to local Lorentz rotations), and one has to recall that $\bar R$ and $\bar\lag^M$ depend on the background tetrad components only through the background metric.  

Continuing the variation inside (\ref{dyn.2.kappa+}) we rewrite it as
\bea
\lag^{dyn}_2(\varkappa)
=&-&
\frac{1}{4\k}\varkappa^{a}{}_{\mu}\frac{\partial \mathcal{K}^{\alpha\beta}}{\partial \bar e^{a}{}_{\mu}}\,
\Big[\bar R_{\alpha\beta}-\k\Big(\bar T^M_{\alf\beta}- \frac{1}{2} \bar g_{\alf\beta}\bar T^M\ \Big)\Bigr]
-
\frac{1}{4\k\bar e}\l( {\cal K}^{\rho\sig} - \frac{1}{2}\bar g^{\rho\sig}{\cal K}^\pi{}_\pi\r)\,
\frac{\delta}{\delta \bar g^{\rho\sigma}} \Big(
\mathcal{K}^{\alpha\beta}\bar R_{\alpha\beta}\Big)
\Big|_{\mathcal K\ \mathrm{fixed}}\nonumber\\ &+& \frac{1}{4\bar e}\l( {\cal K}^{\rho\sig} - \frac{1}{2}\bar g^{\rho\sig}{\cal K}^\pi{}_\pi\r)\,
\frac{\delta}{\delta \bar g^{\rho\sigma}}
\Big[\mathcal{K}^{\alpha\beta}\Big(\bar T^M_{\alf\beta}- \frac{1}{2} \bar g_{\alf\beta}\bar T^M\ \Big)\Big]
\Big|_{\mathcal K\ \mathrm{fixed}}\,,
\label{split_variation}
\eea
where $\l\{|_{\mathcal K\ \mathrm{fixed}}\r\}$ means that a quantity $ {\cal K}^{\rho\sig}$ under Lagrangian derivative does not depend on $\bar g^{\rho\sigma}$ and its derivatives.
For simplicity (without loss of generality), we set $\lag^M = 0$, concentrating on the Ricci flat background in (\ref{split_variation}): 
\begin{equation}
\lag^{dyn}_2(\varkappa) = -
\frac{1}{4\k}\varkappa^{a}{}_{\mu}\frac{\partial \mathcal{K}^{\alpha\beta}}{\partial \bar e^{a}{}_{\mu}}\,
\bar R_{\alpha\beta}
-
\frac{1}{4\k\bar e}\l( {\cal K}^{\rho\sig} - \frac{1}{2}\bar g^{\rho\sig}{\cal K}^\pi{}_\pi\r)\,
\frac{\delta}{\delta \bar g^{\rho\sigma}} \Big(
\mathcal{K}^{\alpha\beta}\bar R_{\alpha\beta}\Big)
\Big|_{\mathcal K\ \mathrm{fixed}}.
\label{split_variation+}
\end{equation}
Here, the first term is proportional to the background Ricci tensor and disappears for the Ricci flat background. This term does not contribute to the field equations, but it contributes to the symmetric (metric) energy-momentum and, thus, has to be preserved. The second term defines propagating  the gravitational perturbations  on a background geometry.  Following the technique given in \cite{GPP_1984,Petrov_KLT_2017} we directly calculate:
\begin{equation}
\frac{\delta}{\delta \bar g^{\rho\sig}} \Big(
\mathcal{K}^{\alpha\beta}\bar R_{\alpha\beta}\Big)\Big|_{\mathcal K\ \mathrm{fixed}}
=
\frac12\Big(
\bar\nabla_{\mu}\bar\nabla^{\mu} \mathcal K_{\rho\sig}
+ \bar g_{\rho\sig}\,\bar\nabla_{\mu}\bar\nabla_{\nu} \mathcal K^{\mu\nu}
- \bar\nabla_{\mu}\nabla_{\rho} \mathcal K_{\sig}{}^{\mu}
- \bar\nabla_{\mu}\nabla_{\sig} \mathcal K_{\rho}{}^{\mu}
\Big).
\label{GL.form.K}
\end{equation}
Then, 
integrating by parts and
discarding a total divergence, one obtains from (\ref{split_variation+}):
\begin{equation}
\lag^{dyn}_2(\varkappa)= -
\frac{1}{4\k}\varkappa^{a}{}_{\mu}\frac{\partial \mathcal{K}^{\alpha\beta}}{\partial \bar e^{a}{}_{\mu}}\,
\bar R_{\alpha\beta}
+\frac{1}{8\kappa\bar e}\Big(
\bar\nabla_\lambda {\cal K}^{\mu\nu}\,\bar\nabla^\lambda \mathcal K_{\mu\nu}
-2\,\bar\nabla_\mu {\cal K}^{\rho\sig}\,\bar\nabla_\rho \mathcal K_{\sig}{}^{\mu} -\frac12\bar\nabla_\lambda\mathcal K_{\pi}{}^{\pi}\nabla^\lambda\mathcal K_{\tau}{}^{\tau}
\Big).
\label{Lg2++}
\end{equation}
It is interesting to derive (\ref{Lg2++}) in evidently depending on the tetrad perturbations $\varkappa^a{}_\mu$ form:
\bea\label{Lg2.kappa}
     \lag^{dyn}_2(\varkappa)= &-&
     \frac{\bar e}{4\k}\Big\{ \big[(\varkappa^a{}_\mu\bar e_a{}^\mu)^2-(\varkappa^a{}_\nu\bar e_a{}^\mu)(\varkappa^b{}_\mu\bar e_b{}^\nu) \big]\bar g^{\alf\beta} \nonumber\\ &-&
     4(\varkappa^a{}_\mu\bar e_a{}^\mu)(\varkappa^b{}^\alf\bar e_b{}^\beta) + 4 (\varkappa^a{}^\alf\bar e_a{}^\mu)(\varkappa^b{}_\mu\bar e_b{}^\beta)+ 2(\varkappa^a{}_\mu\bar e_a{}^\alf)(\varkappa^b{}^\mu\bar e_b{}^\beta) \Big\}\bar R_{\alf\beta} 
      \nonumber\\
          &+& \frac{\bar e}{4\k}\Big\{\bar\nabla_\alf(\varkappa^a{}_\mu\bar e_a{}^\nu) \bar\nabla^\alf(\varkappa^b{}_\nu\bar e_b{}^\mu)+\bar g^{\mu\nu}\bar g_{\sig\rho}\bar\nabla_\alf(\varkappa^a{}_\mu\bar e_a{}^\sig) \bar\nabla^\alf(\varkappa^b{}_\nu\bar e_b{}^\rho) - 2\bar\nabla_\alf(\varkappa^a{}_\mu\bar e_a{}^\mu) \bar\nabla^\alf(\varkappa^b{}_\nu\bar e_b{}^\nu) \Big.\nonumber\\
     &+& \Big. 2\bar\nabla_\alf(\varkappa^a{}_\mu\bar e_a{}^\mu) \bar\nabla^\nu(\varkappa^b{}_\nu\bar e_b{}^\alf) + 2\bar\nabla^\nu(\varkappa^a{}_\mu\bar e_a{}^\mu) \bar\nabla_\alf(\varkappa^b{}_\nu\bar e_b{}^\alf)- 2\bar\nabla^\mu(\varkappa^a{}_\mu\bar e_a{}^\nu) \bar\nabla_\alf(\varkappa^b{}_\nu\bar e_b{}^\alf) \Big. 
     \nonumber\\
     &-& \Big. \bar g^{\alf\mu}\bar g^{\beta\nu}\bar g_{\rho\sig}\bar\nabla_\alf(\k^a{}_\mu\bar e_a{}^\rho) \bar\nabla_\beta(\varkappa^b{}_\nu\bar e_b{}^\sig) +
     \bar g^{\mu\nu} \bar\nabla_\alf(\varkappa^a{}_\mu\bar e_a{}^\alf) \bar\nabla_\beta(\varkappa^b{}_\nu\bar e_b{}^\beta)\Big\}\,.
\eea

The Lagrangian (\ref{dyn_exp2}) is quadratic in $\varkappa^a{}_\mu$ and $\phi^{{\cal A}'}$ the smallness of which and their derivatives are assumed as $\varkappa^a{}_\mu\sim\phi^{{\cal A}'}\ll O(1)$, if the behavior of the background quantities and their derivatives is assumed as $\sim O(1)$. Setting in (\ref{gauge.kappa.phi}) the behavior $\xi^\alf$ and its derivatives as $ \sim \varkappa^a{}_\mu\sim\phi^{{\cal A}'}\ll O(1)$ one can define gauge transformations for $\varkappa^a{}_\mu$ and $\phi^{{\cal A}'}$ correspondingly to (\ref{gauge.kappa.phi}) as
\bea
\varkappa'^{a}{}_{\mu}&=& \varkappa^{a}{}_{\mu}+\Lix\bar e^{a}{}_{\mu}\,,
\m{gauge.lin.kap}\\
\phi'^{{\cal A}'}&=& \phi^{{\cal A}'} +\Lix\bar\Phi^{{\cal A}'}\,.
\m{gauge.lin.phi}
\eea
Substituting these into (\ref{dyn_exp2}) one obtains (\ref{gauge.LT}) up to a second order
\bea
\lag^{dyn}_2(\bar e,\bar\Phi;\varkappa',\phi')
&=&
\lag^{dyn}_2(\bar e,\bar\Phi;\varkappa,\phi) + \Big(\Lix + \frac{1}{2!}\Lix^2 \Big)\bar\lag + \Lix \lag^1
\nonumber\\
&-&\Big[\big(\Lix + \frac{1}{2!}\Lix^2 \big)\bar e^{a}{}_{\mu}+\Lix\varkappa^{a}{}_{\mu}\Big]
\frac{\delta\bar\lag}{\delta\bar e^{a}{}_{\mu}}\nonumber\\
&-&\Big[\big(\Lix + \frac{1}{2!}\Lix^2 \big)\bar\Phi^{{\cal A}'}+\Lix\phi^{{\cal A}'}\Big]
\frac{\delta\bar\lag}{\delta\bar\Phi^{{\cal A}'}} \,,
\label{gauge.LT2}
\eea
where $\lag^1$ is linear in $\k^a{}_\mu$ and $\phi^{{\cal A}'}$ Lagrangian $\lag$. One can see that in spite of the left hand side of (\ref{gauge.LT2}) is quadratic the right hand side contains the linear terms. This looks as a dis-balance. However, one finds that the linear terms up to a divergence, according to the definition of the first variation, disappear themselves
\be
\Lix\bar\lag - \frac{\delta\bar\lag}{\delta\bar e^{a}{}_{\mu}}\Lix\bar e^{a}{}_{\mu} - \frac{\delta\bar\lag}{\delta\bar\Phi^{{\cal A}'}}\Lix\bar\Phi^{{\cal A}'}=0\,.
\m{gauge.linear}
\ee
Anyway, one can see that $\lag^{dyn}_2$ is gauge invariant up to  a divergence (indeed $\Lix\lag^1 = \di_\alf(\xi^\alf\lag^1 )$ and $\Lix^2\bar\lag = \di_\alf(\xi^\alf\Lix\bar\lag )$), if the background equations (\ref{EofM_bar_T}) hold. 
It is illustrative to give the evident gauge transformation on the basis of (\ref{gauge.LT2}) for  the pure gravitational Lagrangian $\lag^{g}_2(\k)$ defined in (\ref{Lg2.kappa}): 
\bea
\lag^{g}_2(\varkappa') =\lag^{g}_2(\varkappa) +\frac{\bar e}{\kappa} \Big(\Lix\varkappa^{a}{}_{\alf} +  \frac{1}{2!}\Lix^2 \bar e^{a}{}_{\alf}\Big)\Big( \bar R_a{}^\alf -\frac12 \bar e_a{}^\alf \bar R\Big)  + {\rm div}\,.
\label{gauge.L2.Ricci.flat}
\eea
One can see that it is gauge invariant up to a divergence on Ricci flat backgrounds. Recall that here $\lag^{g}_2(\varkappa)$ corresponds to the second item in (\ref{split_variation}) without the third term. But vanishing the last just states the vacuum (Ricci flat) background. 

Later we show that the quadratic Lagrangian \eqref{Lg2++} is invariant with respect to linear Lorentz gauge transformations \eqref{Lorentz_varkappa_linear} if the background equations hold. 


\section{The field equations in field-theoretical TEGR}
\label{EofM_TEGR}
\setcounter{equation}{0}

The equations for perturbations are obtained from the dynamical Lagrangian $\lag^{dyn}$ in (\ref{dyn.lag.kappa.phi}), or more concretely in (\ref{dyn.lag.kappa.phi+}),
by varying it with respect to the dynamical variables following the general construction,
see (\ref{EofM_dyn}). Equivalently, they can be cast into the “linear part = source” form (\ref{EofM_dyn+}). For the choice $Q^{A}=\{e^{a}{}_{\mu},\Phi^{{\cal A}'}\}$ and
$q^{A}=\{\varkappa^{a}{}_{\mu},\phi^{{\cal A}'}\}$ it can be considered  in  the form of separated system
(\ref{G_equations})--(\ref{M_equations}):
\begin{equation}
G^{L}{}_{a}{}^{\mu}+\Phi^{L}{}_{a}{}^{\mu}=2\kappa\,\theta^{tot}{}_{a}{}^{\mu},
\label{GLamu},
\end{equation}
\begin{equation}
\Phi^{L}{}_{{\cal A}'}=\theta^{M}{}_{{\cal A}'},
\label{PhiLamu}
\end{equation}
where the linear operators are defined as in (\ref{E_a})--(\ref{Phi_A}) (now with $g^{\cal A}=e^{a}{}_{\mu}$),
while the sources are defined by the same rule as in (\ref{EMT}) and (\ref{EMT+}), i.e. as variational derivatives of $\lag^{dyn}$
with respect to the background fields.
In particular, the source $\theta^{tot}{}_{a}{}^{\mu}$ is expressed through the total (gravitational plus matter) energy--momentum tensor density
of perturbations in the tetrad presentation.

Now we are concentrating on the gravitational equation equation (\ref{GLamu}), and do not pay opinion to pure matter equation (\ref{PhiLamu}) where we do not concretize the matter sector. By the definitions (\ref{E_a}) and (\ref{Phi_a}) for the theory with Lagrangian (\ref{lag+div}) one has
\bea
G^{L}{}_{a}{}^{\mu}+\Phi^{L}{}_{a}{}^{\mu} & =& \frac{\delta}{\delta\bar e^a{}_\mu}\Big[\varkappa^b{}_\nu \frac{\delta(-\bar e\bar T - 2\kappa\bar\lag^M)}{\delta\bar e^b{}_\nu}- 2\kappa \phi^{{\cal A}'}\frac{\delta\bar\lag^M}{\delta\bar\bar\Phi^{{\cal A}'} } \Big]
\nonumber\\
&=&\label{GLamu+}
\frac{\delta}{\delta\bar e^a{}_\mu}\Big[{\cal K}^{\alf\beta}\Big(\frac{\delta (\sqrt{-\bar g}\bar R)}{\delta\bar\gog^{\alf\beta}} - 2\kappa\frac{\delta\bar\lag^M}{\delta\bar\gog^{\alf\beta}}\Big)- 2\kappa \phi^{{\cal A}'}\frac{\delta\bar\lag^M}{\delta\bar\bar\Phi^{{\cal A}'} } \Big]\,,
\eea
where ${\cal K}^{\alf\beta}$ is defined in (\ref{K_def}). Then, taking into account the background equations (\ref{EofM_bar_T}) and using the expression (\ref{GL.form.K}) we rewrite (\ref{GLamu+}) separately for $G^{L}{}_{a}{}^{\mu}$ and $\Phi^{L}{}_{a}{}^{\mu}$:
\begin{equation}
G^{L}{}_{a}{}^{\mu}
=
\frac12\frac{\di\bar g^{\rho\sigma}}{\di\bar e^a{}_\mu }\Big(
\bar\nabla_{\mu}\bar\nabla^{\mu} \mathcal K_{\rho\sig}
+ \bar g_{\rho\sig}\,\bar\nabla_{\mu}\bar\nabla_{\nu} \mathcal K^{\mu\nu}
- \bar\nabla_{\mu}\nabla_{\rho} \mathcal K_{\sig}{}^{\mu}
- \bar\nabla_{\mu}\nabla_{\sig} \mathcal K_{\rho}{}^{\mu}
\Big),
\label{GL.form.K+}
\end{equation}
\begin{equation}
\Phi^{L}{}_{a}{}^{\mu} =-2\k \frac{\di\bar g^{\rho\sigma}}{\di\bar e^a{}_\mu }
\frac{\delta}{\delta\bar g^{\rho\sigma}}\Big[\Big({\cal K}^{\alf\beta}\frac{\delta\bar\lag^M}{\delta\bar\gog^{\alf\beta}}\Big)\Big|_{\mathcal K\ \mathrm{fixed}}+ \phi^{{\cal A}'}\frac{\delta\bar\lag^M}{\delta\bar\bar\Phi^{{\cal A}'} }\Big] \,,
\label{PhiL.form.K+}
\end{equation}
where, by the barred relation (\ref{g_munu}),
\begin{equation}
\frac{\di\bar g^{\rho\sigma}}{\di\bar e^a{}_\mu } = -\l(\bar g^{\mu\rho}\bar e_a{}^\sigma +\bar g^{\mu\sigma}\bar e_a{}^\rho\r)\,.
\label{g.to.e}
\end{equation}
Thus, the expression (\ref{GL.form.K+}) is the result of varying the Lagrangian
\begin{equation}\label{dyn.a}
     (-2\k)\lag^{dyn}_2(\varkappa) = \frac{1}{2!}\varkappa^a{}_\mu \frac{\delta }{\delta\bar e^a{}_\mu} \varkappa^b{}_\nu\frac{\delta }{\delta\bar e^b{}_\nu}\bar e\bar R
     \,
\end{equation}
with respect to $\varkappa^a{}_\mu$:
\begin{equation}\label{dyn.b}
 G^{L}{}_{a}{}^{\mu}
=    (-2\k)\frac{\delta}{\delta\varkappa^a{}_\mu }\lag^{dyn}_2(\k) =  \frac{\delta }{\delta\bar e^a{}_\mu} \varkappa^b{}_\nu\frac{\delta }{\delta\bar e^b{}_\nu}\bar e\bar R
     \,
\end{equation}
and the redefinition from $\varkappa^a{}_\mu$ to ${\cal K}^{\alf\beta}$ using (\ref{K_def}). Equivalently, the expression (\ref{GL.form.K+}) can be obtained by the following way. Let us rewrite  the Lagrangian
(\ref{Lg2++}), taking into account $\bar R_{\alpha\beta}=0$, as
\begin{equation}
(-2\kappa)\lag^{dyn}_2({\cal K})= -\frac{1}{4\bar e}\Big(
\bar\nabla_\lambda {\cal K}^{\mu\nu}\,\bar\nabla^\lambda \mathcal K_{\mu\nu}
-2\,\bar\nabla_\mu {\cal K}^{\rho\sig}\,\bar\nabla_\rho \mathcal K_{\sig}{}^{\mu} -\frac12\bar\nabla_\lambda\mathcal K_{\pi}{}^{\pi}\nabla^\lambda\mathcal K_{\tau}{}^{\tau}
\Big).
\label{dyn.c}
\end{equation}
Then, varying it with respect to ${\cal K}^{\alf\beta}$ directly
and taking into account (\ref{K_def}) one obtains
\begin{equation}\label{dyn.d}
 G^{L}{}_{a}{}^{\mu}
=    (-2\k)\frac{\di {\cal K}^{\alf
\beta}}{\di\varkappa^a{}_\mu}\frac{\delta}{\delta{\cal K}^{\alf
\beta}}\lag^{dyn}_2({\cal K}) 
     \,.
\end{equation}
The source in (\ref{GLamu}) is defined following the general rule (\ref{EMT}) with the dynamic Lagrangian (\ref{dyn.lag.kappa.phi}) or (\ref{dyn.lag.kappa.phi+}) is
\begin{equation}
\theta^{tot}{}_{a}{}^{\mu}=\frac{\delta\lag^{dyn}}{\delta \bar e^a{}_\mu} = 
\frac{\di\bar g^{\rho\sigma}}{\di\bar e^a{}_\mu } \frac{\delta\lag^{dyn}}{\delta \bar g^{\rho\sigma}} =  \frac12
\frac{\di\bar g^{\rho\sigma}}{\di\bar e^a{}_\mu } t^{tot}_{{\rho\sigma}}\,.
\label{theta.tot.amu}
\end{equation}
Here, it was used the definition of the energy-momentum tensor density for perturbations in (\ref{EMT++}).

Let us discuss a gauge invariance of the field equations.
Correspondingly to (\ref{gauge_EofM}), for which $q^A=\{\varkappa^a{}_\mu,\phi^{{\cal A}'} \}$, the equations (\ref{GLamu}), -- with (\ref{GL.form.K+})-(\ref{theta.tot.amu}), -- and (\ref{PhiLamu}) are gauge invariant with respect to transformations (\ref{gauge.kappa.phi}) in the case if the background equations (\ref{EofM_bar_T}) hold and if the field equations (\ref{GLamu}) and (\ref{PhiLamu}) hold themselves. It is important also to check the invariance of the field equations with respect to Lorentz gauge transformations (\ref{gauge.Lorentz}). Keeping in mind the representation of the field equations in the form (\ref{EofM_and_bar}) and using the property (\ref{lag.tot.Lorentz}). As a result, we conclude that 
\begin{equation}
\frac{\delta\lag^{dyn}(\bar e,\bar\Phi,\varkappa^*,\phi)}{\delta\varkappa^*{}^a{}_\mu} = \frac{\delta\lag^{dyn}(\bar e,\bar\Phi,\varkappa,\phi)}{\delta\varkappa{}^a{}_\mu}\,.
\label{eqs.Lorentz.gauge}
\end{equation}
Thus, the field equations are invariant with respect to transformations (\ref{gauge.Lorentz}) without any conditions or restrictions. Finally, one can be easily convinced that transformations (\ref{gauge.kappa.phi}) and (\ref{gauge.Lorentz}) commute as applied to the field equations without any restrictions as well.

One can see that the expressions (\ref{GL.form.K+}),  (\ref{PhiL.form.K+}) and  (\ref{theta.tot.amu}) have the unique coefficient ${\di\bar g^{\rho\sigma}}/{\di\bar e^a{}_\mu }$ defined in (\ref{g.to.e}) and which is not zero by a regularity assumption. Therefore, it can be suppressed in (\ref{GLamu}): 
\begin{equation}
G^{L}_{\rho\sigma}+\Phi^{L}_{\rho\sigma}=\kappa\,t^{tot}_{\rho\sigma},
\label{GLrhosig}
\end{equation}
where
\begin{equation}
G^{L}_{\rho\sigma}
=
\frac12\Big(
\bar\nabla_{\mu}\bar\nabla^{\mu} \mathcal K_{\rho\sig}
+ \bar g_{\rho\sig}\,\bar\nabla_{\mu}\bar\nabla_{\nu} \mathcal K^{\mu\nu}
- \bar\nabla_{\mu}\nabla_{\rho} \mathcal K_{\sig}{}^{\mu}
- \bar\nabla_{\mu}\nabla_{\sig} \mathcal K_{\rho}{}^{\mu}
\Big),
\label{GL.form.K++}
\end{equation}
\begin{equation}
\Phi^{L}_{\rho\sigma} =-2\k 
\frac{\delta}{\delta\bar g^{\rho\sigma}}\Big[\Big({\cal K}^{\alf\beta}\frac{\delta\bar\lag^M}{\delta\bar\gog^{\alf\beta}}\Big)\Big|_{\mathcal K\ \mathrm{fixed}}+ \phi^{{\cal A}'}\frac{\delta\bar\lag^M}{\delta\bar\bar\Phi^{{\cal A}'} }\Big] \,.
\label{PhiL.form.K++}
\end{equation}
In other words, following the logic of constructions in (\ref{G_eqs_munu}) equation  (\ref{GLrhosig}) can be obtained by a direct variation of 
the dynamical Lagrangian $\lag^{dyn}$ in (\ref{dyn.lag.kappa.phi}), or more concretely in (\ref{dyn.lag.kappa.phi+}), with  respect to $\bar g^{\rho\sig}$.

We are more interested in gravitational perturbations. Therefore, for simplicity, we consider equation (\ref{GLrhosig}) for a Ricci-flat background $\bar R = 0$ with $\lag^M=0$ and without dynamic matter fields:
\begin{equation}
G^{L}_{\rho\sigma}(\varkappa)=\kappa\,t^{g}_{\rho\sigma}(\varkappa)\,,
\label{GLrhosig_flat}
\end{equation}
compare with (\ref{G_eqs_munu_g}).

Both of the equations (\ref{G_eqs_munu}) derived from \eqref{b-b11} and (\ref{GLrhosig}) are equivalent to the Einstein equations in the usual form. Thus, they are equivalent between themselves. The same is related to the equations without matter (\ref{G_eqs_munu_g}) and  (\ref{GLrhosig_flat}). Moreover,
one can see that the  expressions for $G^L_{\mu\nu}(\goh)$ in (\ref{b-b21}) and $G^L_{\rho\sig}(\varkappa)$ in (\ref{GL.form.K++}) are quite similar. It is instructive to compare them, comparing $\goh^{\alf\beta}$ defined in (\ref{gog_dec}) with  ${\cal K}^{\alf\beta}$ defined in (\ref{K_def}). We need to express both $\goh^{\alf\beta}$ and ${\cal K}^{\alf\beta}$ through $\varkappa^a{}_\mu$. Already, ${\cal K}^{\alf\beta}$ is presented in (\ref{K_def}), which is  linear in $\varkappa^a{}_\mu$.  By the relation (\ref{g_munu}) one can connect the perturbations $h_{\mu\nu} =g_{\mu\nu}-\bar g_{\mu\nu} $ with $\varkappa^a{}_\mu$:
\begin{equation}
h_{\mu\nu}(\bar e,\varkappa) = \eta_{ab}\big(\bar e^a{}_\mu \varkappa^b{}_\nu + \bar e^b{}_\nu \varkappa^a{}_\mu + \varkappa^a{}_\mu \varkappa^b{}_\nu \big) = h^L_{\mu\nu}(\varkappa) + h^Q_{\mu\nu}(\varkappa^2).
\label{h_kappa}
\end{equation}
By the algebraic relation $\gog^{\alf\beta} = \gog^{\alf\beta}(g_{\mu\nu})$ one has  $\gog^{\alf\beta} = \gog^{\alf\beta}(\bar g_{\mu\nu} + h_{\mu\nu})$. Then 
\begin{eqnarray}
\goh^{\alf\beta} &=& \gog^{\alf\beta}(\bar g_{\mu\nu} + h_{\mu\nu})-\bar\gog^{\alf\beta}(\bar g_{\mu\nu})\nonumber\\ &= &\sum^\infty_{k=1}\frac{1}{k!}\frac{\di^k\bar\gog^{\alf\beta}}{\di\bar g_{\mu\nu}\ldots\di\bar g_{\rho\sig}} \underbrace{h_{\mu\nu}\ldots h_{\rho\sig}}_k = \frac{\di\bar\gog^{\alf\beta}}{\di\bar g_{\mu\nu}} h_{\mu\nu} + \frac{1}{2!}\frac{\di^2\bar\gog^{\alf\beta}}{\di\bar g_{\mu\nu}\di\bar g_{\rho\sig}} {h_{\mu\nu} h_{\rho\sig}} + O(h^3) 
\,.
\label{goh_h}
\end{eqnarray}
On the other hand, ${\cal K}^{\alf\beta}$ in (\ref{K_def}) is rewritten in the form:
\begin{equation}
{\cal K}^{\alf\beta} = \frac{\di\bar\gog^{\alf\beta}}{\di\bar g_{\mu\nu}} h^L_{\mu\nu} = \goh_L^{\alf\beta}\,,
\label{K_lin}
\end{equation}
where $\goh_L^{\alf\beta}$ is linear $\goh^{\alf\beta}$ in $\varkappa^a{}_\mu$.   Thus, $\goh^{\alf\beta}$ differs from ${\cal K}^{\alf\beta}$ in  quadratic and more terms in $\varkappa^a{}_\mu$ 
\begin{eqnarray}
\Delta\goh^{\alf\beta}(\bar e,\varkappa) &=& \goh^{\alf\beta}(\bar e,\varkappa)- {\cal K}^{\alf\beta}(\bar e,\varkappa) = \frac{\di\bar\gog^{\alf\beta}}{\di\bar g_{\mu\nu}} h^Q_{\mu\nu} + \frac{1}{2!}\frac{\di^2\bar\gog^{\alf\beta}}{\di\bar g_{\mu\nu}\di\bar g_{\rho\sig}} {h^L_{\mu\nu} h^L_{\rho\sig}} + O(\varkappa^3)\nonumber\\
&=&
\frac12\bar e \Big\{\big[(\bar e^a{}_\rho\varkappa_a{}^\rho)^2 -(\bar e^a{}_\rho\varkappa_a{}^\sig)(\bar e^b{}_\sig\varkappa_b{}^\rho) ]\bar g^{\alf\beta} - 2(\bar e^b{}_\rho\varkappa_b{}^\rho)\big[(\bar e^a{}^\alf\varkappa_a{}^\beta)+(\bar e^a{}^\beta\varkappa_a{}^\alf)]\nonumber\\
&+&
2(\bar e^a{}^\alf\varkappa_a{}^\sig)(\bar e^b{}^\beta\varkappa_b{}_\sig)+ 2(\bar e^a{}^\alf\varkappa_a{}^\sig)(\bar e^b{}_\sig\varkappa_b{}^\beta)+ 2(\bar e^a{}^\beta\varkappa_a{}^\sig)(\bar e^b{}_\sig\varkappa_b{}^\alf)
\Big\} +  O(\varkappa^3)
\,.
\label{goh_K}
\end{eqnarray}

Recall that equations (\ref{G_eqs_munu_g}) and (\ref{GLrhosig_flat}) are equivalent to the same Einstein equations in the usual form. This means that the difference between the energy-momenta in (\ref{G_eqs_munu_g}) and (\ref{GLrhosig_flat}) is defined by the difference (\ref{goh_K}):
\begin{equation}
\Delta t^{g}_{\mu\nu}(\varkappa) =t^{g}_{\mu\nu}(\goh(\varkappa)) -t^{g}_{\mu\nu}(\varkappa) = \frac{1}{\k} G^{L}_{\mu\nu}(\Delta\goh^{\alf\beta}(\varkappa)).
\label{Delta_tg}
\end{equation}
It can be seen that it is not less than quadratic in $\varkappa$. 

To obtain an explicit expression for $t^{g}_{\mu\nu}(\varkappa)$ in (\ref{GLrhosig_flat}) up to a second order, one can vary (\ref{Lg2++}) directly as in  (\ref{EMT++}) or  (\ref{b-b24}) with respect to a background metric. Note that  variation of ${\cal K}^{\alpha\beta}$ is not provided because it gives items proportional to linear equations in $\varkappa^a{}_\mu$ equations, which have to be hold for quadratic $t^{g}_{\mu\nu}(\varkappa)$. Recall also that the term in (\ref{Lg2++}) proportional $\bar R_{\alf\beta}$ also has to be varied. Another way to obtain an explicit quadratic in $\varkappa$ expression for $t^{g}_{\mu\nu}(\varkappa)$ is to use (\ref{Delta_tg})
\begin{equation}
t^{g}_{\mu\nu}(\varkappa) =t^{g}_{\mu\nu}(\goh(\varkappa)) -\frac{1}{\k} G^{L}_{\mu\nu}(\Delta\goh^{\alf\beta}(\varkappa))
\label{Delta_tg+}
\end{equation}
with taking into account (\ref{goh_K}) and quadratic approximation of (\ref{b-b24})-(\ref{b-b13}) where for expansions one has to use (\ref{h_kappa})-(\ref{goh_h}).


\section{Conserved quantities}
\label{Conserved_quantities}
\setcounter{equation}{0}

In (\ref{G_eqs_munu}) and in (\ref{GLrhosig}), in the case of a vacuum background, $\Bar R_{\mu\nu}=0$, both $\Phi^L_{\mu\nu}\equiv 0$  and $\bar\nabla^\mu G^L_{\mu\nu} \equiv 0$, and thus $\bar\nabla^\mu t_{\mu\nu}^{tot} = 0$. This gives evident advantages in the construction of conserved quantities. For example, if $\bar \xi^\mu$ is a Killing vector of the background, one constructs a related conserved  current, which is the vector density 
\begin{equation}
\bar J^\mu \equiv t_{\mu\nu}^{tot}\bar \xi^\nu :=\bar\nabla_\mu\bar J^\mu=\di_\mu\bar J^\mu=0 \,.
\label{CL_t_current}
\end{equation}
If the background is not vacuum, one has $\Phi^L_{\mu\nu}\neq 0$, $\bar\nabla^\mu G^L_{\mu\nu} \neq 0$ and $\bar\nabla^\mu \big(G^L_{\mu\nu} +\Phi^L_{\mu\nu}  \big)\neq 0$. Thus, there is no conservation of the total energy-momentum $\bar\nabla_\mu t_{\mu\nu}^{tot}\neq 0$. In this case, it is necessary to construct conserved currents and related superpotentials independently of the behavior of $t_{\mu\nu}^{tot}$ on an arbitrary curved background. Here, in the field-theoretical presentation of TEGR we  provide this below.

To obtain such conserved currents and related superpotentials, one has to apply the Noether theorem to the dynamic Lagrangian (\ref{dyn.lag.kappa.phi}), the same (\ref{dyn.lag.kappa.phi+}). But such a procedure is very cumbersome. However, there is a possibility to significantly simplify a situation, and we demonstrate this now. Thus, let us consider the pure gravitational linear item in (\ref{dyn.lag.kappa.phi+}):
\begin{equation}
\lag^G_1=\frac{1}{2\kappa}\varkappa^a{}_\mu\frac{\delta(\bar e\bar T)}{\delta\bar e^a{}_\mu} = -\frac{1}{2\kappa}{\cal K}^{\alf\beta}\frac{\delta(\sqrt{-g}\bar R)}{\delta\bar \gog^{\alf\beta}} =-\frac{1}{2\kappa}{\cal K}^{\alf\beta}\bar R_{\alf\beta}\,,
\label{lag_G_lin}
\end{equation}
where ${\cal K}^{\alf\beta} = \varkappa^a{}_\mu\di\bar \gog^{\alf\beta}/\di\bar e^a{}_\mu$ is defined concretely in (\ref{K_def}). Because $\lag^G_1$ is a scalar density, the Noether 
procedure can be applied directly deriving the identity: 
\begin{equation}
 {\Lix} \lag^G_{1} -\di_\mu \big(\xi^\mu \lag^G_{1}\big)
\equiv 0\,.
 \label{lag_G1_ident}
\end{equation}
It is not
degenerate because $\lag^G_{1}$  in (\ref{lag_G_lin}) contains derivatives of the background metric up to the second order. 

Regrouping (\ref{lag_G1_ident}) one obtains (see Appendix \ref{app:derivation_current})  the identical  conservation of the current $j^\mu$
\begin{equation}
\bar\nabla_\mu j^\mu \equiv \di_\mu j^\mu \equiv 0\,,\qquad j^\mu \equiv  {1 \over {\k}} \Big(G^{L\mu}_{\nu}({\cal K})\xi^\nu +  {\cal K}^{\mu\lam} \Bar  R_{\lam\nu}\xi^\nu + {\cal Z}^\mu(\xi) \Big)\, ,
 \label{identity_div}
\end{equation}
 where ${\cal K}^{\alf\beta}$ is  defined in (\ref{K_def}), $G^{L}_{\mu\nu}({\cal K})$
 is  defined in (\ref{GL.form.K++}) and
\begin{equation}
{\cal Z}^\mu(\xi)  \equiv \l(\zeta^{\rho\sig}\bar\nabla_\rho {\cal K}^\mu_\sig - {\cal K}^{\rho\sig}\bar\nabla_\rho \zeta^\mu_\sig\r) -\half \l(\zeta_{\rho\sig}\bar\nabla^\mu {\cal K}^{\rho\sig} - {\cal K}^{\rho\sig}\bar\nabla^\mu
\zeta_{\rho\sig}\r)
 +\half\l({\cal K}^{\mu\nu}\bar\nabla_\nu \zeta^\rho_\rho - \zeta^\rho_\rho\bar\nabla_\nu {\cal K}^{\mu\nu}\r)
 \m{Zmu}
\end{equation}
 with $\zeta_{\rho\sig} = \half\Lix\bar g_{\rho\sig} =\bar\nabla_{(\rho}\xi_{\rho)}$. Thus, if $\xi^\mu=\bar\xi^\mu$ is a Killing vector of the background, then
$\zeta_{\mu\nu}=\bar\nabla_{(\mu}\bar\xi_{\nu)}=0$, and therefore
\begin{equation}
{\cal Z}^\mu(\bar\xi)=0.
\label{app:Zkill}
\end{equation}
In this case, the identically conserved current \eqref{app:finalcurrent} simplifies to
\begin{equation}
j^\mu\Big|_{\xi=\bar\xi}
=
\frac{1}{\kappa}
\Big(
G^{L\mu}{}_\nu({\cal K})\,\bar\xi^\nu
+{\cal K}^{\mu\lam}\Bar R_{\lam\nu}\bar\xi^\nu
\Big),
\label{app:currentKilling}
\end{equation}
which is the natural intermediate form before using the field equations and passing
to the physically conserved current \eqref{J_current} below.
 
Next, because the current $j^\mu$ is conserved identically (\ref{identity_div}) it has to be represented in the form 
\begin{equation}
j^\mu \equiv \bar\nabla_\nu {\cal J}^{\mu\nu}\equiv \di_\nu {\cal J}^{\mu\nu}\,.
 \label{j=diJ}
\end{equation}
 Indeed, the Klein-Noether identities \cite{Petrov_KLT_2017} constructed for the Lagrangian (\ref{lag_G_lin}) allow us to represent the current $j^\mu $ in the form (\ref{j=diJ}), where superpotential is presented by the antisymmetric tensor density 
\begin{equation}
 {\cal J}^{\mu\nu} \equiv 
 {1 \over \k} \l({\cal K}^{\rho[\mu}\bar\nabla_\rho\xi^{\nu]}+
\xi^{[\mu}\bar\nabla_\sig {\cal K}^{\nu]\sig}-\bar\nabla^{[\mu}{\cal K}^{\nu
]}_\sig \xi^\sig \r). 
\label{alaAbbottDeser}
\end{equation}
Another direct way of a transformation from $j^\mu$ to ${\cal J}^{\mu\nu}$ is given in Appendix \ref{app:superpotential_construction}. We note the famous properties of the superpotential \eqref{alaAbbottDeser}: first, it has been constructed without the smallness assumption for ${\cal K}^{\alf\beta}$; second, it is linear in ${\cal K}^{\alf\beta}$ and its first derivatives only; third, it holds in an arbitrary curved background spacetime. At last, formally a superpotential of such a form has been constructed in \cite{AbbottDeser_1982} in GR on the anti-de Sitter background with the Killing vectors. 

However, the conservation laws (\ref{identity_div}) and (\ref{j=diJ}) are identities only. To obtain physically sensible conservation laws, one has to use field equations. In our case, we use (\ref{GLrhosig}) and rewrite the current in (\ref{identity_div}) as a conserved current
\begin{equation}
{\cal J}^\mu = {1 \over {\k}} \Big[ \l(\k t^{tot\,\mu}_\nu -\Phi^{L\,\mu}_{\nu} +  {\cal K}^{\mu\lam} \Bar  R_{\lam\nu}\r)\xi^\nu + {\cal Z}^\mu(\xi) \Big] := \bar\nabla_\mu {\cal J}^\mu=\di_\mu {\cal J}^\mu=0 \, ,
 \label{J_current}
\end{equation}
The role of interaction with an arbitrary curved background is played by the term
$  ({\cal K}^{\mu\lam} \Bar  R_{\lam\nu}  -\Phi^{L\,\mu}_{\nu})\xi^\nu$. In the case of Ricci-flat background and with the Killing vectors of a background the conserved current in (\ref{J_current}) transforms to the current in (\ref{CL_t_current}). Finally, the identity  (\ref{j=diJ}) transforms to physically sensible conservation law
\begin{equation}
{\cal J}^\mu = \bar\nabla_\nu {\cal J}^{\mu\nu} = \di_\nu {\cal J}^{\mu\nu}\,.
 \label{J=diJ}
\end{equation}
 Thus, by a construction of (\ref{alaAbbottDeser})-(\ref{J=diJ}) we have resolved a construction of conserved quantities for perturbations on arbitrary curved backgrounds. 

By the differential conservation laws (\ref{J_current}) and (\ref{J=diJ}) one can define integral conserved quantities in the standard way. Integrating through 4-volume the equality (\ref{J_current}) one obtains a conserved (in time with related boundary conditions) quantity  ${\cal P}(\xi)$ placed on 3-dimensional section $\Sigma =: x^0 = t = {\rm const}$:
\begin{equation}\label{ICQJa}
    {\cal P}(\xi) = \int_\Sigma d^3x {\cal J}^{0}(\xi)\,,
\end{equation}
where the ``0''- component is related to a time coordinate, whereas Latin indexes from the middle of alphabet, for example, ``i'', are related to space coordinates.
 The quantity (\ref{ICQJa}) with the use of (\ref{J=diJ}) is reduced to a surface integral that is called the Noether charge:
\begin{equation}\label{ICQJab}
   {\cal P}(\xi) =  \oint_{\di\Sigma} ds_i {\cal J}^{0i}(\xi)\,,
\end{equation}
where $\di\Sigma$ is a boundary of $\Sigma$, and $ds_i $ is an integration element on  $\di\Sigma$. By construction, the quantity ${\cal P}(\xi)$ in both  (\ref{ICQJa}) and  (\ref{ICQJab}) is  a scalar; indeed, it is invariant both with respect to coordinate transformations and with respect to local Lorentz rotations.

\section{Applications}
\label{Applications}
\setcounter{equation}{0}

\subsection{Linear equations and gravitational waves}

In this subsection, we consider the linear tetrad perturbation theory. For simplicity, we  study it on a Ricci-flat background without matter perturbations. Thus, instead of (\ref{GLrhosig}) or (\ref{GLrhosig_flat}) we consider the equation
\begin{equation}
\label{eq:K-operator-original}
G^L_{\rho\sigma}(\mathcal{K})
=
\frac12\big(\bar\nabla_{\lambda}\bar\nabla^{\lambda}\mathcal{K}_{\rho\sigma}
+ \bar g_{\rho\sigma}\,\bar\nabla_{\lambda}\bar\nabla_{\tau}\mathcal{K}^{\lambda\tau}
- \bar\nabla_{\lambda}\bar\nabla_{\sigma}\mathcal{K}_{\rho}{}^{\lambda}
- \bar\nabla_{\lambda}\bar\nabla_{\rho}\mathcal{K}_{\sigma}{}^{\lambda}\big) = 0 .
\end{equation}
Let us check a gauge invariance  of this equation with respect to gauge transformation (\ref{gauge.lin.kap}). As before, we set the behavior of $\xi^\alf$ and its derivatives as $ \sim \varkappa^a{}_\mu\sim\phi^{{\cal A}'}\ll O(1)$ with their derivatives. Then, substituting (\ref{gauge.lin.kap}) in the definition (\ref{K_def}) of $\mathcal{K}^{\alf\beta}$ we obtain
\begin{equation}
\label{eq:K-gauge}
\mathcal{K}'^{\alf\beta} =  \mathcal{K}^{\alf\beta}  -\bar e\big(\bar\nabla^\alf\xi^\beta + \bar\nabla^\beta\xi^\alf - \bar g^{\alf\beta}\bar\nabla_\rho\xi^\rho\big) = \mathcal{K}^{\alf\beta} +\pounds_\xi\bar\gog^{\alf\beta}\,.
\end{equation}
Substitution of this transformation into  (\ref{eq:K-operator-original}) gives
\begin{equation}
\label{eq:GLK-gauge}
G^L_{\rho\sigma}(\mathcal{K}')
=
G^L_{\rho\sigma}(\mathcal{K}) + 
\big(\delta^\pi_\rho \delta^\tau_\sig -\half \bar g^{\pi\tau}\bar g_{\rho\sig}\big)\pounds_\xi \bar R_{\pi\tau} .
\end{equation}
The derivation of this formula is completely analogous to the derivation of Eq. (2.2.81) in the book \cite{Petrov_KLT_2017}.
Thus, the linear equations (\ref{eq:K-operator-original}) are gauge invariant with respect to transformations (\ref{eq:K-gauge}), the same as (\ref{gauge.lin.kap}), on the Ricci-flat background. In the case of non-Ricci-flat background one needs to check the gauge invariance of the linear equation equation $G^L_{\rho\sigma} + \Phi^L_{\rho\sigma} = 0$.

Such a gauge invariance allows us to cancel non-physical degrees of freedom. Let us rewrite (\ref{eq:K-operator-original}) replacing covariant derivatives in the form 
\begin{equation}
\label{eq:K-operator-reordered}
\begin{aligned}
G^L_{\rho\sigma}(\mathcal{K}) 
=&\;
\frac12\Big(\bar\nabla_{\lambda}\bar\nabla^{\lambda}\mathcal{K}_{\rho\sigma}
+ \bar g_{\rho\sigma}\,\bar\nabla_{\lambda}\bar\nabla_{\tau}\mathcal{K}^{\lambda\tau}
- \bar\nabla_{\sigma}\bar\nabla_{\lambda}\mathcal{K}_{\rho}{}^{\lambda}
- \bar\nabla_{\rho}\bar\nabla_{\lambda}\mathcal{K}_{\sigma}{}^{\lambda}
\\[0.3em]
&\;
- \bar R_{\lambda\sigma}\,\mathcal{K}_{\rho}{}^{\lambda}
- \bar R_{\lambda\rho}\,\mathcal{K}_{\sigma}{}^{\lambda}
- \bar R_{\rho}{}^{\tau}{}_{\lambda\sigma}\,\mathcal{K}_{\tau}{}^{\lambda}
- \bar R_{\sigma}{}^{\tau}{}_{\lambda\rho}\,\mathcal{K}_{\tau}{}^{\lambda}\Big) = 0.
\end{aligned}
\end{equation}
Formally it is the Lichnerowicz equation \cite{Lichnerowicz_1961} applied to $\mathcal{K}^{\rho\sigma}$.
Now, we impose a transversality condition on $\mathcal{K}_{\mu\nu}$ (the Lorentz condition): 
\begin{equation}
\bar\nabla_{\tau}\mathcal{K}^{\lambda\tau}=0,
\label{eq:K-transverse}
\end{equation}
restricting 4 degrees of freedom, i.e.\ $\mathcal{K}_{\mu\nu}$ becomes divergence--free. Then, for the Ricci-flat background equation (\ref{eq:K-operator-reordered}) transforms to 
\begin{equation}
\label{eq:variation-Ricci-flat-transverse}
G^L_{\rho\sigma}(\mathcal{K}) 
=
\half
\bar\Box\mathcal{K}_{\rho\sigma}- \bar R_{\rho\tau\lambda\sigma}\,\mathcal{K}^{\tau\lambda}
= 0\,,
\end{equation}
where we use the convenient notation $\bar\Box\equiv\bar\nabla_{\lambda}\bar\nabla^{\lambda}$. Now, it is important to define residual degrees of freedom. Substitution of (\ref{eq:K-gauge}) into (\ref{eq:K-transverse}) and (\ref{eq:variation-Ricci-flat-transverse}) and keeping in mind $\bar R_{\alf\beta}=0$ gives, respectively,
\begin{equation}
\bar\Box\xi^\alf =0,
\label{eq:K-transverse-xi}
\end{equation}
\begin{equation}
\bar\nabla_\rho\bar\Box\xi_\sig + \bar\nabla_\sig\bar\Box\xi_\rho - \bar g_{\rho\sig} \bar\nabla_\alf\bar\Box\xi^\alf =0.
\label{eq:GL-transverse-xi}
\end{equation}
One can see that these restrictions on $\xi^\alf$ are quite consistent. Moreover, condition (\ref{eq:K-transverse-xi}) is enough to satisfy (\ref{eq:GL-transverse-xi}). Thus, using $\xi^\alf$ satisfying (\ref{eq:K-transverse-xi}) one can restrict other 4 degrees of freedom. Among them, one can cancel the trace
\begin{equation}
\mathcal{K}^{\alf}{}_{\alf} = 0
\label{eq:K-trace-zero}
\end{equation}
that corresponds to $\bar\nabla_\alf\xi^\alf = 0$. Indeed, it is consistent with the trace of the equality (\ref{eq:GL-transverse-xi}): $\bar\nabla_\alf\bar\Box\xi^\alf =\bar\Box\bar\nabla_\alf\xi^\alf =0$, where it was used $\bar R_{\alf\beta}=0$ again.

Despite the fact that the quantity $\mathcal{K}^{\rho\sigma}$ defined in (\ref{K_def}) is not directly related to metric perturbations, equation (\ref{eq:variation-Ricci-flat-transverse}), being the Lichnerowicz wave equation, conditions (\ref{eq:K-transverse}) and (\ref{eq:K-trace-zero}) and restrictions (\ref{eq:K-transverse-xi}) and (\ref{eq:GL-transverse-xi}) quite correspond to those relations for metric perturbations in GR. On the other hand, initially the tetrad perturbations $\varkappa^a{}_\mu$ have 16 components, whereas $\mathcal{K}^{\rho\sigma}$ have 10 components. The obvious reason why such a transformation is permissible consists in the fact that tetrad components are included (up to a divergence) in the TEGR Lagrangian (\ref{lag+div}) only through the metric (\ref{g_munu}). 

It is instructive to consider this question explicitly in detail. Recall that the field equations in the field-theoretical TEGR are invariant (\ref{eqs.Lorentz.gauge}) with respect to Lorentz gauge transformations (\ref{gauge.Lorentz}). Thus, the linear equations (\ref{eq:K-operator-original}) have to be invariant with respect to linear version of (\ref{gauge.Lorentz}):
\begin{equation}
\varkappa^*{}^a{}_\mu = \varkappa^a{}_\mu + \epsilon^a{}_b\bar e^b{}_\mu\,,
\label{Lorentz_varkappa_linear}
\end{equation}
where $\epsilon^a{}_b=\epsilon^a{}_b(x)$ is a matrix of a small local Lorentz rotation such as $\epsilon^a{}_b\sim\xi^\alf\sim \varkappa^a{}_\mu\ll 1$ with the same relations for their derivatives, note also $\epsilon_{ab} = -\epsilon_{ba}$. Concerning (\ref{eq:K-operator-original}), first let us analyze a transformation of $\mathcal{K}_{\alf\beta}$ defined in (\ref{K_def}) under transformation (\ref{Lorentz_varkappa_linear}). By the antisymmetry $\epsilon_{ab} = -\epsilon_{ba}$ one has
\begin{equation}
\mathcal{K}^*_{\alf\beta} = \mathcal{K}_{\alf\beta}\,.
\label{K*=K}
\end{equation}
This means that the linear equations (\ref{eq:K-operator-original}) in whole are invariant with respect to the transformation (\ref{Lorentz_varkappa_linear}). One has to recall that in the linear version gauge transformations (\ref{Lorentz_varkappa_linear}) and (\ref{gauge.lin.kap}) commute.

It is often convenient to convert the Lorentz index to a spacetime index
using the background tetrad,
\begin{equation}
\varkappa_{\mu\nu}\equiv \bar e_{a\mu}\,\varkappa^{a}{}_{\nu}\,.
\label{kap.munu=kap.amu}
\end{equation}
Then we have the irreducible decomposition
\begin{equation}
\varkappa_{\mu\nu}=\varkappa_{(\mu\nu)}+\varkappa_{[\mu\nu]},
\qquad
\#\big(\varkappa_{(\mu\nu)}\big)=10,
\qquad
\#\big(\varkappa_{[\mu\nu]}\big)=6\,.
\label{irreducible.dec}
\end{equation}
Under the transformation (\ref{Lorentz_varkappa_linear}) one has 
\begin{equation}
\varkappa_{(\mu\nu)}^*=\varkappa_{(\mu\nu)},
\qquad
\varkappa^*_{[\mu\nu]}= \varkappa_{[\mu\nu]}+\epsilon_{\mu\nu},
\label{[munu].(munu)}
\end{equation}
so the antisymmetric part \(\varkappa_{[\mu\nu]}\) is \emph{pure Lorentz gauge}
and can be set to zero by a suitable choice of \(\epsilon_{\mu\nu}\).
After this Lorentz gauge fixing one is left with
\begin{equation}
\#\big(\varkappa_{(\mu\nu)}\big)=10,
\qquad
\#\big(\varkappa_{[\mu\nu]}\big)=0\,.
\label{irreducible.dec}
\end{equation}
It is important to make a remark. If one goes beyond TEGR to theories where local Lorentz symmetry is broken,
e.g. some modified teleparallel models, the antisymmetric sector can become
dynamical and the number of propagating degrees of freedom may increase.

Next, bringing into mind the definition of  $\mathcal{K}_{\alf\beta}$ in (\ref{K_def}) with obvious and simple algebraic dependence on $\varkappa_{(\mu\nu)}$ we conclude that the variables $\mathcal{K}_{\alf\beta}$ in (\ref{eq:variation-Ricci-flat-transverse}) can be equivalently exchanged by $\varkappa_{(\mu\nu)}$. 
Therefore, instead of the wave-type equation (\ref{eq:variation-Ricci-flat-transverse}) one can consider the equation
\begin{equation}
\bar\Box\varkappa_{(\rho\sig)}
-
2\bar R_{\rho}{}^{\tau\lam}{}_{\sigma}\,
\varkappa_{(\tau\lam)}
=0
\label{eq:kappa_wave_Lorentz_invariant}
\end{equation}
that is invariant under local Lorentz gauge transformations (\ref{Lorentz_varkappa_linear}).
Equivalently, one has the conditions 
\begin{equation}
\bar\nabla^{\mu}\varkappa_{(\mu\nu)}=0
\qquad \varkappa^\alf{}_\alf = 0
\label{two.conditions}
\end{equation}
instead of (\ref{eq:K-transverse}) and (\ref{eq:K-trace-zero}) leading to  a transverse-traceless-type gauge on suitable backgrounds
so that only two independent polarizations for the $\varkappa_{(\mu\nu)}$ components remain. Consideration of weak gravitational waves in TEGR with the use of (\ref{eq:kappa_wave_Lorentz_invariant}) looks preferable because $\varkappa_{(\mu\nu)}$ is directly and simply connected with the tetrad perturbations, see (\ref{kap.munu=kap.amu}).

Let us illustrate \eqref{eq:kappa_wave_Lorentz_invariant} and \eqref{two.conditions} on the example of a flat background in the Lorentzian coordinates $\{t,x,y,z\}$ equivalent to  $\{0,1,2,3\}$. Then, equation \eqref{eq:kappa_wave_Lorentz_invariant} becomes a simple wave equation 
\begin{equation}
\bar\Box\varkappa_{(\rho\sig)} 
=0
\label{eq:Box_wave_Lorentz}
\end{equation}
and the background tetrad acquires the simplest form
\begin{equation}\label{bare_Lorentz}
\bar e^a {}_\mu = {\rm diag}\left(1,1,1,1
\right).
\end{equation}
Setting a flat wave propagation in $x$-direction and achieving the TT-gauge we obtain
\begin{equation}\label{varkappa_TT}
\varkappa^{TT}_{(\mu\nu)}(t-x) = \left(
\begin{array}{cccc}
 {0} & 0 & 0 & 0 \\
 0 & {0} & 0 & 0 \\
 0 & 0 & {\varkappa_{22}} & \varkappa_{23} \\
 0 & 0 & \varkappa_{32} & {\varkappa_{33}} \\
\end{array}
\right)
\end{equation}
with $\varkappa_{33} = -\varkappa_{22}$ and $\varkappa_{23} =\varkappa_{32}$. It is important to clarify the correspondence between \eqref{varkappa_TT} and the linear perturbations of the tetrad components $\varkappa^a{}_\mu$. Let us rewrite the definition of $\varkappa_{(\mu\nu)}$ explicitly:
\begin{equation}\label{varkappa()}
\varkappa_{(\mu\nu)} = \frac12\eta_{ab}\big(\bar e^a {}_\mu\varkappa^b{}_{\nu} + \bar e^a {}_\nu\varkappa^b{}_{\mu}\big).
\end{equation}
Then, keeping in mind \eqref{bare_Lorentz} one has nonzero $\varkappa^a{}_\mu$ in the TT-gauge
\begin{equation}\label{varkappa_TT_amu}
\varkappa^{aTT}_{~~\mu} = \left(
\begin{array}{cccc}
 {0} & \varkappa^0{}_1 & \varkappa^0{}_2 & \varkappa^0{}_3 \\
 \varkappa^0{}_1 & {0} & \varkappa^1{}_2 & \varkappa^1{}_3 \\
 \varkappa^0{}_2 & -\varkappa^1{}_2 & {\varkappa^2{}_{2}} & \varkappa^2{}_3 \\
 \varkappa^0{}_3 & -\varkappa^1{}_3 & \varkappa^3{}_2 & -\varkappa^2{}_{2} \\
\end{array}
\right).
\end{equation}
Note that only the component $\varkappa^2{}_{2}$ and the sum $\varkappa^2{}_3 + \varkappa^3{}_2$ must have a wave behavior with the dependence on the  coordinate $(t-x)$ that follows from \eqref{varkappa_TT}. Other components of \eqref{varkappa_TT_amu} can have an arbitrary dependence without destruction of \eqref{varkappa_TT}. The wave solution, exactly as \eqref{varkappa_TT_amu}, was obtained in \cite{Farrugia_2018} within the framework of $f(T)$ theory. It was stressed that the difference with the related picture in TEGR does not appear in linear order. Thus, our result \eqref{varkappa_TT_amu} obtained within the field-theoretical approach in TEGR supports this conclusion.

Finally, it is instructive to compare the wave solution \eqref{varkappa_TT} with the related solution in TT-gauge in GR. Let us rewrite the relation \eqref{h_kappa} in the form:
\begin{equation}\label{h_varkappa}
h_{\mu\nu} = 2\varkappa_{(\mu\nu)} + O(\varkappa^2).
\end{equation}
Then rewriting the TT-gauge  $h^{TT}_{\mu\nu}$ in the usual form 
\begin{equation}\label{h_TT}
h^{TT}_{\mu\nu}(t-x) = \left(
\begin{array}{cccc}
 {0} & 0 & 0 & 0 \\
 0 & {0} & 0 & 0 \\
 0 & 0 & {h_{22}} & h_{23} \\
 0 & 0 & h_{32} & {h_{33}} \\
\end{array}
\right)\,
\end{equation}
and noting the factor 2 in \eqref{h_varkappa}, we find the difference from \eqref{varkappa_TT} only by  the quadratic order $O(\varkappa^2)$.

\subsection{Black-hole charges: computational examples}

Here, we illustrate using formulae for conserved charges given in section \ref{Conserved_quantities} on the examples of black-hole solutions as asymptotically flat ones with the choice of flat background.
In the Ricci-flat case and for a background Killing vector
$\bar\xi^\mu$, the current $\mathcal{J}^{\mu}$  in \eqref{J_current} reduces to the usual conserved vacuum current,
and the Noether charge \eqref{ICQJab} is computed from the superpotential
$\mathcal{J}^{\mu\nu}$ in \eqref{alaAbbottDeser}. Since for the black-hole
examples below the integration surface is the sphere $r=\mathrm{const}$ at
spatial infinity, we use a spherical coordinates $(t,r,\theta,\phi) = (0,1,2,3)$ in the flat background. Then, only the $(01)$-component of $\mathcal{J}^{0i}$ contributes:  
\begin{equation}
\mathcal{P}(\bar\xi)
=
\lim_{r\to\infty}
\oint_{\partial\Sigma_r} ds_i\,\mathcal{J}^{0i}(\bar\xi)
=
\lim_{r\to\infty}
\int \mathcal{J}^{01}(\bar\xi)\,d\theta\,d\phi .
\label{eq:BH_charge_surface}
\end{equation}    
We apply \eqref{eq:BH_charge_surface} first to the Schwarzschild geometry and
then to the Kerr geometry.

\subsubsection{Schwarzschild solution}

In Schwarzschild coordinates $(t,r,\theta,\phi)$ the metric in the
$(-+++)$ convention is
\begin{equation}
ds^2
=
-\left(1-\frac{2M}{r}\right)dt^2
+\frac{dr^2}{1-\frac{2M}{r}}
+r^2 d\theta^2
+r^2\sin^2\theta\,d\phi^2 ,
\label{eq:Schw_metric_cov}
\end{equation}
whereas the flat background metric in the same coordinates is
\begin{equation}
d\bar s^2
=
-dt^2+dr^2+r^2 d\theta^2+r^2\sin^2\theta\,d\phi^2 .
\label{eq:Schw_background_cov}
\end{equation}

A more natural choice of the tetrad for the metric \eqref{eq:Schw_metric_cov} is the diagonal tetrad: 
\begin{equation}
e^a{}_\mu = {\rm diag}\l\{\sqrt{1-{2M}/{r}},\,\frac{1}{\sqrt{1-{2M}/{r}}},\,r,\,r\sin\theta \r\}
\label{eq:Schw_tetrad}
\end{equation}
with background tetrad
\begin{equation}
\bar e^a{}_\mu = {\rm diag}\{1,\,1,\,r,\,r\sin\theta \}
\label{eq:Schw_background_tetrad}
\end{equation}
Hence, the non-vanishing tetrad perturbations $\varkappa^a{}_\mu=e^a{}_\mu-\bar e^a{}_\mu$
are
\begin{equation}
\varkappa^0{}_t=\sqrt{1-\frac{2M}{r}}-1,
\qquad
\varkappa^1{}_r=\frac{1}{\sqrt{1-{2M}/{r}}}-1.
\label{eq:Schw_tetrad_pert}
\end{equation}
Using the definition \eqref{K_def}, one finds that the only non-zero components
of $\mathcal{K}^{\mu\nu}$ are
\begin{align}
\mathcal{K}^{00}
&=
-\frac{2Mr\sin\theta}{\sqrt{1-\frac{2M}{r}}},
\nonumber\\
\mathcal{K}^{11}
&=
-\frac{2Mr\sin\theta}{\sqrt{1-\frac{2M}{r}}},
\nonumber\\
\mathcal{K}^{22}
&=
\frac{\sin\theta\left(\sqrt{1-\frac{2M}{r}}-1\right)^2}
{\sqrt{1-\frac{2M}{r}}},
\nonumber\\
\mathcal{K}^{33}
&=
\frac{\csc\theta\left(\sqrt{1-\frac{2M}{r}}-1\right)^2}
{\sqrt{1-\frac{2M}{r}}}.
\label{eq:Schw_K_components}
\end{align}

To obtain the mass of the object one has to calculate the $01$-component of the superpotential \eqref{alaAbbottDeser} with the components 
\eqref{eq:Schw_K_components} and with the timelike Killing vector:
\begin{equation}
\bar\xi^\mu_{(t)}=(-1,0,0,0).
\label{eq:Schw_time_Killing}
\end{equation}
As a result, 
\begin{equation}
{\cal J}^{01} = \frac{r \sin \theta  \left(\frac{1}{\sqrt{1-{2 M}/{r}}}-1\right)}{8 \pi } \underset{r\rightarrow\infty}{\sim} \frac{M\sin\theta}{4\pi}
\label{eq:J_01}
\end{equation}
where we also include the asymptotics at ${r\rightarrow\infty}$. 
After substituting this 
into \eqref{eq:BH_charge_surface}, one obtains 
\begin{equation}
\mathcal{P}\big(\bar\xi_{(t)}\big)=M.
\label{eq:Schw_mass_charge}
\end{equation}

Let us discuss the quite acceptable result \eqref{eq:Schw_mass_charge}.
The non-vanishing covariant metric perturbation components as the difference of \eqref{eq:Schw_metric_cov} and \eqref{eq:Schw_background_cov} $h_{\mu\nu}=g_{\mu\nu}-\bar g_{\mu\nu}$ and their asymptotics at ${r\rightarrow\infty}$ are 
\begin{equation}
h_{tt}=\frac{2M}{r} \underset{r\rightarrow\infty}{\sim} \frac{2M}{r},
\qquad
h_{rr}=\frac{1}{1-{2M}/{r}}-1 \underset{r\rightarrow\infty}{\sim} \frac{2M}{r}.
\label{eq:Schw_h_cov}
\end{equation}
Note that just the behavior as in \eqref{eq:Schw_h_cov} allows us to obtain the acceptable value $M$ for mass of the Schwarzschild black hole in many known approaches for calculating the mass in an asymptotically flat spacetime in GR (see, for example, \cite{GPP_1984,Petrov_1995,Petrov_1997} and the book \cite{Petrov_KLT_2017} with numerous references therein). 

Now, with the use of \eqref{eq:Schw_background_tetrad} and \eqref{eq:Schw_tetrad_pert} we calculate $h^L_{\mu\nu}(\varkappa)$ defined in \eqref{h_kappa}:
\begin{equation}
h^L_{tt}(\varkappa)=-2\sqrt{1-\frac{2M}{r}}+2 \underset{r\rightarrow\infty}{\sim} \frac{2M}{r},
\qquad
h^L_{rr}(\varkappa)=\frac{2}{\sqrt{1-{2M}/{r}}}-2 \underset{r\rightarrow\infty}{\sim} \frac{2M}{r}.
\label{eq:Schw_hL_cov}
\end{equation}
It can be seen  that the linear in  $\varkappa^a{}_\mu$ the quantity ${\cal K}^{\alf\beta}$ in \eqref{K_lin} with \eqref{eq:Schw_hL_cov} and the  linear $\goh^{\alf\beta}_L$ defined in \eqref{goh_h} with \eqref{eq:Schw_h_cov} have the same behavior at $r\rightarrow\infty$. Thus, the result  \eqref{eq:Schw_mass_charge} is not surprising.

It is instructive to illustrate the situation with such a calculation on the example of the Schwarzschild solution in isotropic Cartesian
coordinates $(t,x,y,z)=(0,1,2,3)$:
\begin{equation}
ds^2
=
-\left(\frac{1-M/2r}{1+M/2r}\right)^2 dt^2
+
\left(1+\frac{M}{2r}\right)^4
\left(dx^2+dy^2+dz^2\right),
\label{eq:Schw_iso_metric}
\end{equation}
where $r\equiv\sqrt{x^2+y^2+z^2}$ differs from $r$ introduced in \eqref{eq:Schw_metric_cov} and \eqref{eq:Schw_background_cov}.
The background Minkowski metric is chosen as
\begin{equation}
d\bar s^2=-dt^2+dx^2+dy^2+dz^2 .
\label{eq:Schw_iso_background}
\end{equation}
A convenient diagonal tetrad is
\begin{equation}
e^a{}_\mu =\left(1+\frac{M}{2r}\right)^2{\rm diag}\Big\{\frac{1-M/2r}{(1+M/2r)^3},\,1,\,1,\,1 \Big\},
\label{eq:Schw_iso_tetrad}
\end{equation}
while the background tetrad is chosen as 
\begin{equation}
\bar e^a{}_\mu ={\rm diag}\{1,\,1,\,1,\,1 \}\,.
\label{eq:Schw_Min_tetrad}
\end{equation}
Then, non-zero components of the tetrad perturbations are
\begin{equation}
\varkappa^0{}_0=\frac{1-M/2r}{1+M/2r} -1 ,
\qquad
\varkappa^1{}_1=\varkappa^2{}_2=\varkappa^3{}_3=
\left(1+\frac{M}{2r}\right)^2 -1,
\label{eq:Schw_pert_iso_tetrad}
\end{equation}
Then the only 
non-zero components of $\mathcal{K}^{\mu\nu}$ are
\begin{align}
\mathcal{K}^{00}
&=
-\frac{M}{2r}\,\frac{8+9M/2r+ 3(M/2r)^2}
{1+M/2r},
\nonumber\\
\mathcal{K}^{11}
&=
\mathcal{K}^{22}
=
\mathcal{K}^{33}
= \l(\frac{M}{2r} \r)^2\, \frac{3+M/2r}{1+M/2r}.
\label{eq:Schw_iso_K_components}
\end{align}

To use the formula \eqref{eq:BH_charge_surface} for calculating the mass one has to transform the Cartesian isotropic coordinates $(t,x,y,z)$ to spherical coordinates. Due to the covariance of the formalism in general, such  transformations are allowed and we do it.
Then, the charge \eqref{eq:BH_charge_surface} with the superpotential \eqref{alaAbbottDeser} computed for \eqref{eq:Schw_iso_K_components} in spherical form and with the Killing vector \eqref{eq:Schw_time_Killing}  is again
\begin{equation}
\mathcal{P}\big(\bar\xi_{(t)}\big)=M.
\label{eq:Schw_iso_mass_charge}
\end{equation}

Thus, both the Schwarzschild-coordinate and the isotropic-coordinate realizations
lead to the same black-hole mass. Let us compare them. Using \eqref{eq:Schw_Min_tetrad} and \eqref{eq:Schw_pert_iso_tetrad} we calculate $h^L_{\mu\nu}(\varkappa)$ defined in \eqref{h_kappa}:
\begin{equation}
h^L_{00}(\varkappa)
=-2\frac{1-M/2r}{1+M/2r} -2 \underset{r\rightarrow\infty}{\sim} \frac{2M}{r},
\qquad
h^L_{11}(\varkappa)=h^L_{22}(\varkappa)=h^L_{33}(\varkappa)=2
\left(1+\frac{M}{2r}\right)^2 -2 \underset{r\rightarrow\infty}{\sim} \frac{2M}{r}.
\label{eq:Schw_hL_iso_tetrad}
\end{equation}
For a comparison it is necessary to transform \eqref{eq:Schw_hL_cov} to Cartesian coordinates conserving the main order only contributing to mass:
\begin{equation}
h^L_{00}(\varkappa)=\underset{r\rightarrow\infty}{\sim} \frac{2M}{r},
\qquad
h^L_{ij}(\varkappa)=\underset{r\rightarrow\infty}{\sim} \frac{2M}{r}\frac{x^ix^j}{r^2}.
\label{eq:Schw_hL_cov-dec}
\end{equation}
Both \eqref{eq:Schw_hL_iso_tetrad} and \eqref{eq:Schw_hL_cov-dec} have the unique behavior ${\sim} 1/r$. In fact, it is the simplest definition of asymptotically flat spacetime. If such a behavior corresponds to different definitions of perturbations, but is related to the same isolated system, its charges (for example, mass) are the same \cite{Petrov_1995,Petrov_1997}. Therefore, the coincidence of \eqref{eq:Schw_mass_charge} and  \eqref{eq:Schw_iso_mass_charge} is the inevitable result.

In \cite{EPT_2025}, we develop the correct definitions of conserved quantities in TEGR using the classical Noether theorem without introducing background quantities. These quantities, being Lorentz covariant/invariant, depend essentially on inertial spin connections.  Among other applications, this formalism has been applied to calculate the mass for the Schwarzschild black hole \cite{EPT19,EPT_2020,EKPT_2021,EKPT_2021a}. By this calculation, it is quite crucial  to find for a chosen tetrad a relative inertial spin connection. It is provided by a ``turning of gravity'' procedure and does not lead  to a uniquely acceptable and general result.  This ambiguity is connected directly with the problem of non-localization of energy in GR (here in TEGR), and can be bypassed for concrete solutions under consideration, not generally.  

Unlike the above, ambiguity of non-localization of energy in the field-theoretical formalism is directly connected with  a choice of background structures. It is specifically connected with a concrete solution. Thus, (as an example) in the field-theoretical TEGR to obtain acceptable charges for an isolated system with flat background one needs to find a tetrad and its perturbations that lead, at least, to  behavior  $h^L_{\mu\nu}(\varkappa)={\sim} 1/{r}$.

\subsubsection{Kerr solution}

In Boyer--Lindquist coordinates $(t,r,\theta,\phi)$ we write the Kerr metric as
\begin{align}
ds^2
={}&
-\left(1-\frac{2Mr}{\rho^2}\right)dt^2
-\frac{4aMr\sin^2\theta}{\rho^2}\,dt\,d\phi
+\frac{\rho^2}{\Delta}\,dr^2
+\rho^2\,d\theta^2
\nonumber\\
&\hspace{2cm}
+\sin^2\theta
\left(
r^2+a^2+\frac{2a^2Mr\sin^2\theta}{\rho^2}
\right)d\phi^2 ,
\label{eq:Kerr_metric_cov}
\end{align}
where
\begin{equation}
\rho^2=r^2+a^2\cos^2\theta,
\qquad
\Delta=r^2-2Mr+a^2 .
\label{eq:Kerr_rho_Delta}
\end{equation}
The background metric is chosen again in the
form \eqref{eq:Schw_background_cov}, then the components of the background tetrad is chosen in the form \eqref{eq:Schw_background_tetrad}, respectively. 
The relevant Killing vectors of the background for calculating mass and angular momentum (timelike and rotation ones) are
\begin{equation}
\bar\xi^\mu_{(t)}=(-1,0,0,0),
\qquad
\bar\xi^\mu_{(\phi)}=(0,0,0,1).
\label{eq:Kerr_Killing_vectors}
\end{equation}
A physical tetrad related to the solution \eqref{eq:Kerr_metric_cov} we chose in two different representations.
The first tetrad is the one corresponding to Eq.~(3.7) of
\cite{EPT:2024wmy}. In component form it can be written as
%
\begin{equation}\label{eq:Kerr_tetrad_I_components}
  e^a{}_\mu = \left(
\begin{array}{cccc}
 {p_0} & 0 & 0 & {p_4} \\
 0 & {p_1} & 0 & 0 \\
 0 & 0 & {p_2} & 0 \\
 0 & 0 & 0 & {p_3} \\
\end{array}
\right).
\end{equation}
with
\begin{equation}
p_0=\frac{\sqrt{\rho^2-2Mr}}{\rho},
\qquad
p_1=\frac{\rho}{\sqrt{\Delta}},
\qquad
p_2=\rho,
\qquad
p_3=\frac{\sqrt{\Delta}\,\rho\sin\theta}{\sqrt{\rho^2-2Mr}},
\qquad
p_4=\frac{2aMr\sin^2\theta}{\rho\sqrt{\rho^2-2Mr}}.
\label{eq:Kerr_tetrad_I}
\end{equation}

The second tetrad is the one corresponding to Eq.~(3.20) of
\cite{EPT:2024wmy}. Adapting the notation to the present conventions, we write
its non-zero components as
\begin{equation}\label{eq:Kerr_tetrad_II_components}
    e^a{}_\mu =  \left(
\begin{array}{cccc}
 {q_0} & 0 & 0 & {q_4} \\
 0 & {q_1} & 0 & 0 \\
 0 & 0 & {q_2} & 0 \\
 {q_4} & 0 & 0 & {q_3} \\
\end{array}
\right)
\end{equation}
where
\begin{eqnarray}
  q_0&=&-\frac{|g_{03}|}{g_{03}}\,A(g)\,\bigl[g_{00}+B(g)\bigr],\qquad
q_1=\sqrt{g_{11}},\qquad
q_2=\sqrt{g_{22}},\nonumber\\
q_3&=&\frac{|g_{03}|}{g_{03}}\,A(g)\,\bigl[B(g)-g_{33}\bigr],
\qquad
q_4=|g_{03}|\,A(g),
\label{eq:Kerr_tetrad_II}
\end{eqnarray}
with
\begin{equation}
A(g)=
\sqrt{
\frac{g_{00}-g_{33}-2B(g)}
{(g_{00}+g_{33})^2-4g_{03}^2}
},
\qquad
B(g)=\sqrt{g_{03}^2-g_{00}g_{33}} .
\label{eq:Kerr_tetrad_II_AB}
\end{equation}

Each of tetrads can be expressed in a unified form
\begin{equation}\label{s_tetrad}
    e^a{}_\mu =  \left(
\begin{array}{cccc}
 {s_0} & 0 & 0 & {s_4} \\
 0 & {s_1} & 0 & 0 \\
 0 & 0 & {s_2} & 0 \\
 {s_5} & 0 & 0 & {s_3} \\
\end{array}
\right)\,,
\end{equation}
then the tetrad perturbations are  also expressed in a unified form
\begin{equation}\label{varkappa_tetrad}
\varkappa^a{}_\mu =  \left(
\begin{array}{cccc}
 {s_0}-1 & 0 & 0 & {s_4} \\
 0 & {s_1-1} & 0 & 0 \\
 0 & 0 & {s_2}-r & 0 \\
 {s_5} & 0 & 0 & {s_3}-r\sin\theta \\
\end{array}
\right).
\end{equation}
Next, for \eqref{varkappa_tetrad}, using the definition \eqref{K_def}, we find the only non-zero components of $\mathcal{K}^{\mu\nu}$ also in unified form:
\begin{align}
\mathcal{K}^{00}
&=
\bar e\big(\varkappa^{\hat 0}{}_0 \bar e_{\hat 0}{}^0 - \varkappa^{\hat 1}{}_1 \bar e_{\hat 1}{}^1  - \varkappa^{\hat 2}{}_2 \bar e_{\hat 2}{}^2 -\varkappa^{\hat 3}{}_3 \bar e_{\hat 3}{}^3    \big),
\nonumber\\
\mathcal{K}^{11}
&=
\bar e\big(\varkappa^{\hat 0}{}_0 \bar e_{\hat 0}{}^0 - \varkappa^{\hat 1}{}_1 \bar e_{\hat 1}{}^1  + \varkappa^{\hat 2}{}_2 \bar e_{\hat 2}{}^2 +\varkappa^{\hat 3}{}_3 \bar e_{\hat 3}{}^3    \big),
\nonumber\\
\mathcal{K}^{22}
&=
\bar e\big(\varkappa^{\hat 0}{}_0 \bar e_{\hat 0}{}^0 + \varkappa^{\hat 1}{}_1 \bar e_{\hat 1}{}^1  - \varkappa^{\hat 2}{}_2 \bar e_{\hat 2}{}^2 +\varkappa^{\hat 3}{}_3 \bar e_{\hat 3}{}^3    \big),
\nonumber\\
\mathcal{K}^{33}
&=
\bar e\big(\varkappa^{\hat 0}{}_0 \bar e_{\hat 0}{}^0 + \varkappa^{\hat 1}{}_1 \bar e_{\hat 1}{}^1  + \varkappa^{\hat 2}{}_2 \bar e_{\hat 2}{}^2 -\varkappa^{\hat 3}{}_3 \bar e_{\hat 3}{}^3    \big),
\nonumber\\
\mathcal{K}^{03}
&=
-\bar e\big(\varkappa^{\hat 0}{}_3 \bar e_{\hat 0}{}^0\bar g^{33} + \varkappa^{\hat 3}{}_0 \bar e_{\hat 3}{}^3 \bar g^{00}   \big).
\label{eq:Kerr_K_components}
\end{align}

The 01-component of the superpotential \eqref{alaAbbottDeser} related to the timelike Killing vector and its asymptotics for both chosen tetrads is
\begin{equation}
{\cal J}^{01}(\bar\xi_{(t)})  = \frac{r \left[2  {s_1}\sin \theta-\di_r{s_2}\sin \theta-\di_r{s_3}\right]}{
   8 \pi } \underset{r\rightarrow\infty}{\sim} \frac{M \sin \theta}{4 \pi }\,.
\end{equation}
Analogously, the 01-component of the superpotential \eqref{alaAbbottDeser} related to the rotation Killing vector and its asymptotics for both chosen tetrads is
\begin{equation}
     {\cal J}^{01}(\bar\xi_{(\phi)})  = \frac{r\sin\theta \left[2s_4 -r  {s_5}\sin \theta-r\di_r{s_4}+r^2\di_r{s_5}\sin\theta\right]}{
   16\pi } \underset{r\rightarrow\infty}{\sim} \frac{3 a M \sin ^3\theta}{8 \pi }\,.
\end{equation}
Thus, for both tetrads \eqref{eq:Kerr_tetrad_I_components} and \eqref{eq:Kerr_tetrad_II_components}, the explicit evaluation of the
Noether charges in the present setup gives
\begin{equation}
\mathcal{P}\big(\bar\xi_{(t)}\big)=M,
\qquad
\mathcal{P}\big(\bar\xi_{(\phi)}\big)=aM\,,
\label{eq:Kerr_tetrad_result}
\end{equation}
which are quite acceptable values.

To conclude the subsection, let us briefly discuss the results. 

First, 
in the direct metric-density calculation, the only non-vanishing components of the perturbation are the diagonal pieces together with the mixed $03-$ and
$30$-components. Their behavior at $r\rightarrow\infty$ just corresponds to the asymptotically flat spacetime and also gives the result \eqref{eq:Kerr_tetrad_result}, see \cite{Petrov_1995,Petrov_1997}. The behavior of the components \eqref{eq:Kerr_K_components} (we do not derive them here because they are very cumbersome)  corresponds to the behavior of metric perturbations that also leads to \eqref{eq:Kerr_tetrad_result}. 

Second, it is instructive to compare the calculation of charges for the Kerr solution here, for both tetrads \eqref{eq:Kerr_tetrad_I_components} and \eqref{eq:Kerr_tetrad_II_components}, with the charges obtained for the Kerr solution in \cite{EPT:2024wmy} with the same tetrads also. In \cite{EPT:2024wmy}, for the calculations of charges, one uses the classical Noether conserved quantities where crucial quantities are pairs of tetrad + inertial spin connection. The last is determined by the ``turning of gravity'' procedure, and background is not used.  By this way, it turns out that the tetrad \eqref{eq:Kerr_tetrad_I_components} does not allow us to find a related inertial spin connection, giving a wrong result $\mathcal{P}\big(\bar\xi_{(\phi)}\big)=aM/3$. Here, in the field theoretical approach, a crucial factor is the asymptotics of ${\cal K}^{\alpha\beta}$. Recall that the asymptotics is defined by a choice of a background structure, and it turns out the same for both tetrads, giving the same result.

\section{Concluding remarks}
\label{Concluding_remarks}
\setcounter{equation}{0}

In the present paper, we have represented TEGR in a field-theoretical form, where tetrad and matter perturbations are classified as dynamical fields. All the properties of the formalism developed in the paper present a novelty in the framework of TEGR. The related dynamic Lagrangian for the perturbations is constructed, variation of which with respect to dynamic variables leads to the equations for the perturbations. These constructions are quite general. 

The formalism can be applied to an arbitrary solution (model under consideration). A background can be chosen in an arbitrary way, usually corresponding to the chosen solution of TEGR.  One of the important properties is that perturbations and all related expressions are exact, not infinitesimal, or approximated. The next approximations can easily be provided. As an application, we have considered linear gravitational equations on the Ricci-flat background and analyzed their properties, finally deriving gravitational wave equations and the tetrad perturbations in TT-gauge.

By the Noether theorem, we have constructed the conserved currents, superpotentials and charges for the perturbations. As applications, by new formulae, we have calculated the mass for the Schwarzschild and Kerr black holes and the angular momentum for the Kerr solution. The results are acceptable, which signals that the new formalism is quite powerful and can be used further. 

We have included an important consideration related to the definition of gauge (inner) transformations and the study of invariance with respect to them. We have introduced the gauge transformations induced by both the general covariance and invariance with respect to local Lorentz rotations. The first kind can be constructed for each field theory if it is generally covariant, thus, it is very well known in the framework of GR. On the other hand, the Lorentz gauge transformations  are defined first and related invariance of the dynamical Lagrangian and field equations are studied first.

We suppose that the present results are the first step in studying  teleparallel theories in the framework of the field-theoretical approach. We plan to develop it, for example, for $f(T)$-theory and other generalizations of TEGR. It is clear that such a study must be significantly cumbersome. However, principal differences are also expected. In TEGR we had the possibility to ignore inertial spin connection because it is not a dynamic field. This significantly simplifies the study. In more general theories, it is not so because components of the inertial spin connection can be dynamic variables. Then, the background components of the inertial spin connection have to satisfy the background equations, and the perturbations of the inertial spin connection have to be the dynamic fields on the total list of the dynamic fields of the field-theoretical presentation of the theory.

Consideration of the cosmological perturbations in the framework of the field-theoretical reformulation of GR, has been quite successful \cite{KP_2013,KP_2014}. 
Therefore, among possible future applications of the field-theoretical approach to teleparallel theories suggested here, we consider the study of cosmological perturbations.

\bigskip

\noindent {\bf Acknowledgments. } EE has been supported in part by the ministry of absorption, the “Program of Support of
High Energy Physics” Grant by Israeli Council for Higher Education and by the Israel Science Fund (ISF) grant
No. 1698/22”, the study by AP was conducted under the state assignment of Lomonosov Moscow State University.

\bigskip

\noindent {\bf Data Availability Statement.} This manuscript has no associated data. [Authors’ comment: All calculations have been presented directly in the text.]

\appendix

\section{A possibility of perturbation of inertial spin connection in TEGR}
\setcounter{equation}{0}
\label{Perturbation_ISC_TEGR}

In this Appendix, we comment more explicitly the possible perturbation of the inertial
spin connection in the field-theoretical reformulation of TEGR remarked in subsection \ref{Dynamic_Lagrangian_TEGR+}. Now, we note that the notions of ${\omega}{}^a{}_{b\mu}$, $\lag^{dyn}_2$, $G^L_{\alpha\beta}(\varkappa)$, $\epsilon^a{}_b$ have been introduced in  \eqref{ISCdef}, \eqref{Lg2.kappa}, \eqref{GL.form.K++}, \eqref{Lorentz_varkappa_linear}, respectively. Thus, 
one could formally extend the field-theoretical decomposition \eqref{decomposition.TEGR}
by writing
\begin{equation}
{\omega}{}^a{}_{b\mu}
=
\bar{\omega}{}^a{}_{b\mu}
+
\alpha^a{}_{b\mu},
\qquad
\alpha_{ab\mu}=-\alpha_{ba\mu}.
\label{spin.connection.decomposition}
\end{equation}
However, in TEGR this does not introduce an additional dynamical sector.
Indeed, the spin connection is inertial and flat. Therefore the perturbation
$\alpha^a{}_{b\mu}$ is restricted by the condition that the perturbed
teleparallel connection remains flat. At the linear level this gives\footnote{
Here $\bar D_\mu$ denotes the Lorentz-covariant derivative constructed with the
background inertial spin connection $\bar\omega{}^a{}_{b\mu}$. For a Lorentz-algebra-valued
quantity $\epsilon^a{}_b$ one has
\[
\bar D_\mu\epsilon^a{}_b
=
\partial_\mu\epsilon^a{}_b
+
\bar\omega{}^a{}_{c\mu}\epsilon^c{}_b
-
\bar\omega{}^c{}_{b\mu}\epsilon^a{}_c .
\]
Similarly,
\[
\bar D_\mu\alpha^a{}_{b\nu}
=
\partial_\mu\alpha^a{}_{b\nu}
+
\bar\omega{}^a{}_{c\mu}\alpha^c{}_{b\nu}
-
\bar\omega{}^c{}_{b\mu}\alpha^a{}_{c\nu}.
\]
The spacetime index $\nu$ is regarded here as the one-form index of the
connection perturbation; equivalently, if the Levi-Civita connection is also
included, its contribution drops out after antisymmetrization in $\mu,\nu$.
Thus
\[
\bar D_{[\mu}\alpha^a{}_{|b|\nu]}
=
\frac12\left(
\bar D_\mu\alpha^a{}_{b\nu}
-
\bar D_\nu\alpha^a{}_{b\mu}
\right).
\]
Expanding the curvature of the perturbed spin connection gives
\[
R^a{}_{b\mu\nu}(\bar\omega+\alpha)
=
\bar R^a{}_{b\mu\nu}
+
2\bar D_{[\mu}\alpha^a{}_{|b|\nu]}
+
O(\alpha^2).
\]
Since the teleparallel background is flat, $\bar R^a{}_{b\mu\nu}=0$, the
linearized flatness condition is precisely
$\bar D_{[\mu}\alpha^a{}_{|b|\nu]}=0$. Locally, by the covariant Poincare lemma
for a flat background connection, this implies
$\alpha^a{}_{b\mu}=\bar D_\mu\epsilon^a{}_b$ with
$\epsilon_{ab}=-\epsilon_{ba}$, up to a possible sign convention for the
infinitesimal Lorentz parameter.
}

\begin{equation}
\bar D_{[\mu}\alpha^a{}_{|b|\nu]}=0,
\label{flatness.alpha.linear}
\end{equation}
where $\bar D_\mu$ is the Lorentz-covariant derivative constructed with
$\bar\omega{}^a{}_{b\mu}$. Locally, this condition implies
\begin{equation}
\alpha^a{}_{b\mu}=\bar D_\mu\epsilon^a{}_b,
\qquad
\epsilon_{ab}=-\epsilon_{ba},
\label{alpha.pure.gauge}
\end{equation}
so that $\alpha^a{}_{b\mu}$ represents the infinitesimal Lorentz-gauge part
of the inertial connection rather than a new propagating field. Moreover,
because of the identity (\ref{lag+div}), the dependence of the TEGR
Lagrangian on the inertial spin connection is contained only in a total
divergence. Consequently, if one keeps the spin-connection perturbation
formally, the dynamical Lagrangian can differ from
(\ref{dyn.lag.kappa.phi}) only by a boundary term,
\begin{equation}
\lag^{dyn}(\bar e,\bar\omega;\varkappa,\alpha;\bar\Phi,\phi)
=
\lag^{dyn}(\bar e,\bar\omega;\varkappa,0;\bar\Phi,\phi)
+
\text{div},
\label{dyn.lag.with.alpha}
\end{equation}
and the Lagrange derivatives with respect to the bulk dynamical variables are
unchanged. Thus, after suppressing total divergences, the quadratic
dynamical Lagrangian and the field equations obtained below are the same as
those obtained with fixed inertial spin connection:
\begin{equation}
\lag^{dyn}_2(\varkappa,\alpha)
=
\lag^{dyn}_2(\varkappa)+\text{div},
\qquad
G^L_{\rho\sigma}(\varkappa,\alpha)
=
G^L_{\rho\sigma}(\varkappa),
\label{alpha.no.bulk.effect}
\end{equation}
 Thus the omission
of independent spin-connection perturbations is not an additional dynamical
assumption in TEGR; it is a consequence of the inertial, flat, pure-gauge
character of ${\omega}{}^a{}_{b\mu}$ and of the boundary-term identity
(\ref{lag+div}). This statement is special to TEGR. In more general
teleparallel theories, such as non-linear $f(T)$ models, the spin-connection
sector cannot in general be removed in the same way.

\section{Derivation of the identically conserved current}
\setcounter{equation}{0}
\label{app:derivation_current}

In this appendix, we present in detail the derivation of the identically conserved current
\eqref{identity_div} from the Noether identity \eqref{lag_G1_ident}.  
Our task is only to show carefully how the three structures
\[
G^{L\mu}{}_{\nu}({\cal K})\xi^\nu,
\qquad
{\cal K}^{\mu\lam}\Bar R_{\lam\nu}\xi^\nu,
\qquad
{\cal Z}^\mu(\xi)
\]
arise and combine into an identically conserved current.

We start from the linear in ${\cal K}^{\alf\beta}$ Lagrangian \eqref{lag_G_lin}, rewriting it here:
\begin{equation}
\lag^G_1
=
-\frac{1}{2\kappa}\,{\cal K}^{\alf\beta}\Bar R_{\alf\beta}\,.
\label{app:L1start}
\end{equation}
Since $\lag^G_1$ is a scalar density of weight
$+1$, its Lie derivative satisfies the identity  \eqref{lag_G1_ident}, and we rewrite it here as well,
\begin{equation}
 {\Lix} \lag^G_{1} -\di_\mu \big(\xi^\mu \lag^G_{1}\big)\equiv 0\,.
\label{app:noeth0}
\end{equation}
The crucial point is that $\lag^G_1$ depends on
the background metric through $\Bar R_{\alf\beta}$, and therefore it contains second
derivatives of the background metric.  For this reason the identity \eqref{app:noeth0}
is not algebraically trivial.

Substituting \eqref{app:L1start} into \eqref{app:noeth0}, we obtain
\begin{equation}
0
\equiv
-\frac{1}{2\kappa}\,{\Lix}\!\left({\cal K}^{\alf\beta}\Bar R_{\alf\beta}\right)
+\frac{1}{2\kappa}\,\di_\mu\!\left(\xi^\mu {\cal K}^{\alf\beta}\Bar R_{\alf\beta}\right).
\label{app:noeth1}
\end{equation}
Using the Leibniz rule, we rewrite this as
\begin{equation}
0
\equiv
-\frac{1}{2\kappa}\,(\Lix {\cal K}^{\alf\beta})\Bar R_{\alf\beta}
-\frac{1}{2\kappa}\,{\cal K}^{\alf\beta}\Lix \Bar R_{\alf\beta}
+\frac{1}{2\kappa}\,\di_\mu\!\left(\xi^\mu {\cal K}^{\alf\beta}\Bar R_{\alf\beta}\right).
\label{app:noeth2}
\end{equation}
We now treat the two Lie derivatives separately. 
First, since ${\cal K}^{\alf\beta}$ is a contravariant tensor density of weight $+1$,
its Lie derivative is
\begin{equation}
\Lix {\cal K}^{\alf\beta}
=
\xi^\rho\bar\nabla_\rho {\cal K}^{\alf\beta}+{\cal K}^{\alf\beta}\bar\nabla_\rho\xi^\rho
-{\cal K}^{\rho\beta}\bar\nabla_\rho \xi^\alf
-{\cal K}^{\alf\rho}\bar\nabla_\rho \xi^\beta
 .
\label{app:LieK}
\end{equation}
Multiplying this by the symmetric tensor $\Bar R_{\alf\beta}$ and using the symmetry
${\cal K}^{\alf\beta}={\cal K}^{\beta\alf}$,  we find
\begin{align}
(\Lix {\cal K}^{\alf\beta})\Bar R_{\alf\beta}
={}&
\xi^\rho(\bar\nabla_\rho {\cal K}^{\alf\beta})\Bar R_{\alf\beta}
-2{\cal K}^{\rho\beta}\Bar R_{\alf\beta}\bar\nabla_\rho\xi^\alf
+{\cal K}^{\alf\beta}\Bar R_{\alf\beta}\bar\nabla_\rho\xi^\rho
\nonumber\\
={}&
\bar\nabla_\rho\!\left(\xi^\rho{\cal K}^{\alf\beta}\Bar R_{\alf\beta}\right)
-\xi^\rho {\cal K}^{\alf\beta}\bar\nabla_\rho \Bar R_{\alf\beta}
-2{\cal K}^{\rho\beta}\Bar R_{\alf\beta}\bar\nabla_\rho\xi^\alf .
\label{app:LieK2}
\end{align}
Substituting \eqref{app:LieK2} into \eqref{app:noeth2}, the total divergence in
\eqref{app:LieK2} cancels the explicit divergence already present in
\eqref{app:noeth2}.  Thus \eqref{app:noeth2} becomes
\begin{equation}
0
\equiv
\frac{1}{2\kappa}\,\xi^\rho {\cal K}^{\alf\beta}\bar\nabla_\rho \Bar R_{\alf\beta}
+\frac{1}{\kappa}\,{\cal K}^{\rho\beta}\Bar R_{\alf\beta}\bar\nabla_\rho\xi^\alf
-\frac{1}{2\kappa}\,{\cal K}^{\alf\beta}\Lix \Bar R_{\alf\beta}.
\label{app:noeth3}
\end{equation}

The last term in \eqref{app:noeth3} is the nontrivial one.  Here one must not use
the final tensorial formula for $\Lix \Bar R_{\alf\beta}$, because that would hide the
second derivatives of the background metric and would not expose the Noether current.
Instead, one has to treat $\Lix \Bar R_{\alf\beta}$ as the variation of the Ricci tensor
induced by the metric variation
\begin{equation}
\delta \bar g_{\mu\nu}=\Lix \bar g_{\mu\nu}=2\zeta_{\mu\nu},
\qquad
\zeta_{\mu\nu}\equiv \bar\nabla_{(\mu}\xi_{\nu)} .
\label{app:zetadef}
\end{equation}
For an arbitrary symmetric variation $\delta \bar g_{\mu\nu}$ one has the standard
formula
\begin{equation}
\delta \Bar R_{\alf\beta}
=
\frac12
\Big(
\bar\nabla_\rho\bar\nabla_\alf \delta \bar g^\rho{}_\beta
+\bar\nabla_\rho\bar\nabla_\beta \delta \bar g^\rho{}_\alf
-\bar\nabla_\alf\bar\nabla_\beta \delta \bar g^\rho{}_\rho
-\bar\nabla_\rho\bar\nabla^\rho \delta \bar g_{\alf\beta}
\Big).
\label{app:deltaRicci}
\end{equation}
After substituting $\delta \bar g_{\mu\nu}=2\zeta_{\mu\nu}$ from \eqref{app:zetadef},
this becomes
\begin{equation}
\Lix \Bar R_{\alf\beta}
=
\bar\nabla_\rho\bar\nabla_\alf \zeta^\rho{}_\beta
+\bar\nabla_\rho\bar\nabla_\beta \zeta^\rho{}_\alf
-\bar\nabla_\alf\bar\nabla_\beta \zeta^\rho{}_\rho
-\bar\nabla_\rho\bar\nabla^\rho \zeta_{\alf\beta}.
\label{app:LieRicciViaZeta}
\end{equation}
We now contract \eqref{app:LieRicciViaZeta} with ${\cal K}^{\alf\beta}$ and exchange the order of derivatives extracting the total derivatives
in such a way that all second derivatives act on ${\cal K}^{\alf\beta}$ rather
than on $\zeta_{\mu\nu}$.  Since ${\cal K}^{\alf\beta}$ is symmetric, the first two terms
in \eqref{app:LieRicciViaZeta} are equal after relabeling the dummy indices.  Therefore,
\begin{align}
-\frac{1}{2}\,{\cal K}^{\alf\beta}\Lix \Bar R_{\alf\beta}
={}&
-{\cal K}^{\alf\beta}\bar\nabla_\rho\bar\nabla_\alf \zeta^\rho{}_\beta
+\frac12 {\cal K}^{\alf\beta}\bar\nabla_\alf\bar\nabla_\beta \zeta^\rho{}_\rho
+\frac12 {\cal K}^{\alf\beta}\bar\nabla_\rho\bar\nabla^\rho \zeta_{\alf\beta}.
\label{app:RicciBlock1}
\end{align}
Now integrate each term by parts once:
\begin{align}
-{\cal K}^{\alf\beta}\bar\nabla_\rho\bar\nabla_\alf \zeta^\rho{}_\beta
={}&
-\bar\nabla_\rho\!\left({\cal K}^{\alf\beta}\bar\nabla_\alf \zeta^\rho{}_\beta\right)
+(\bar\nabla_\rho {\cal K}^{\alf\beta})\bar\nabla_\alf \zeta^\rho{}_\beta ,
\label{app:parts1}
\\[1mm]
\frac12 {\cal K}^{\alf\beta}\bar\nabla_\alf\bar\nabla_\beta \zeta^\rho{}_\rho
={}&
\frac12 \bar\nabla_\alf\!\left({\cal K}^{\alf\beta}\bar\nabla_\beta \zeta^\rho{}_\rho\right)
-\frac12 (\bar\nabla_\alf {\cal K}^{\alf\beta})\bar\nabla_\beta \zeta^\rho{}_\rho ,
\label{app:parts2}
\\[1mm]
\frac12 {\cal K}^{\alf\beta}\bar\nabla_\rho\bar\nabla^\rho \zeta_{\alf\beta}
={}&
\frac12 \bar\nabla_\rho\!\left({\cal K}^{\alf\beta}\bar\nabla^\rho \zeta_{\alf\beta}\right)
-\frac12 (\bar\nabla_\rho {\cal K}^{\alf\beta})\bar\nabla^\rho \zeta_{\alf\beta}.
\label{app:parts3}
\end{align}
At this stage one exchanges the order of derivatives extracting the total derivatives once more, now moving
the derivatives from $\zeta_{\mu\nu}$ onto ${\cal K}^{\alf\beta}$.  Collecting the resulting second-derivative terms one obtains precisely the linear operator
$G^L_{\mu\nu}({\cal K})$ defined in \eqref{GL.form.K++}, while all the first-derivative
boundary terms combine into the divergence of ${\cal Z}^\mu(\xi)$.  The result is
\begin{equation}
-\frac{1}{2\kappa}\,{\cal K}^{\alf\beta}\Lix \Bar R_{\alf\beta}
=
-\frac{1}{\kappa}\,G^L_{\mu\nu}({\cal K})\,\zeta^{\mu\nu}
-\frac{1}{\kappa}\,\di_\mu {\cal Z}^\mu(\xi),
\label{app:mainRicciResult}
\end{equation}
where
\begin{equation}
{\cal Z}^\mu(\xi)
=
\left(
\zeta^{\rho\sig}\bar\nabla_\rho {\cal K}^\mu{}_\sig
-
{\cal K}^{\rho\sig}\bar\nabla_\rho \zeta^\mu{}_\sig
\right)
-\frac12
\left(
\zeta_{\rho\sig}\bar\nabla^\mu {\cal K}^{\rho\sig}
-
{\cal K}^{\rho\sig}\bar\nabla^\mu \zeta_{\rho\sig}
\right)
+\frac12
\left(
{\cal K}^{\mu\nu}\bar\nabla_\nu \zeta^\rho{}_\rho
-
\zeta^\rho{}_\rho\bar\nabla_\nu {\cal K}^{\mu\nu}
\right),
\label{app:Zres}
\end{equation}
which is exactly \eqref{Zmu}.  

Substituting \eqref{app:mainRicciResult} into \eqref{app:noeth3}, we get
\begin{equation}
0
\equiv
-\frac{1}{\kappa}\,G^L_{\mu\nu}({\cal K})\,\zeta^{\mu\nu}
-\frac{1}{\kappa}\,\di_\mu {\cal Z}^\mu(\xi)
+\frac{1}{2\kappa}\,\xi^\rho {\cal K}^{\alf\beta}\bar\nabla_\rho \Bar R_{\alf\beta}
+\frac{1}{\kappa}\,{\cal K}^{\rho\beta}\Bar R_{\alf\beta}\bar\nabla_\rho\xi^\alf .
\label{app:noeth4}
\end{equation}
The first term still contains $\zeta^{\mu\nu}$.  Since $G^L_{\mu\nu}({\cal K})$ is
symmetric, we can write
\begin{equation}
G^L_{\mu\nu}({\cal K})\,\zeta^{\mu\nu}
=
G^L_{\mu\nu}({\cal K})\,\bar\nabla^\mu\xi^\nu
=
\bar\nabla_\mu\!\Big(G^{L\mu}{}_\nu({\cal K})\,\xi^\nu\Big)
-\xi^\nu \bar\nabla_\mu G^{L\mu}{}_\nu({\cal K}).
\label{app:Gzeta}
\end{equation}
Substituting \eqref{app:Gzeta} into \eqref{app:noeth4}, we obtain
\begin{align}
0
\equiv{}&
-\frac{1}{\kappa}\,\di_\mu\!\Big(G^{L\mu}{}_\nu({\cal K})\,\xi^\nu\Big)
-\frac{1}{\kappa}\,\di_\mu {\cal Z}^\mu(\xi)
\nonumber\\
&\quad
+\frac{1}{\kappa}\,\xi^\nu \bar\nabla_\mu G^{L\mu}{}_\nu({\cal K})
+\frac{1}{2\kappa}\,\xi^\rho {\cal K}^{\alf\beta}\bar\nabla_\rho \Bar R_{\alf\beta}
+\frac{1}{\kappa}\,{\cal K}^{\rho\beta}\Bar R_{\alf\beta}\bar\nabla_\rho\xi^\alf .
\label{app:noeth5}
\end{align}
Now we use the generalized Bianchi identity for the operator
$G^L_{\mu\nu}({\cal K})$.  A direct divergence of \eqref{GL.form.K++},
followed by commutation of covariant derivatives, gives
\begin{equation}
\bar\nabla_\mu G^{L\mu}{}_\nu({\cal K})
\equiv
\bar\nabla_\mu\!\Big({\cal K}^{\mu\lam}\Bar R_{\lam\nu}\Big)
-\frac12\,{\cal K}^{\alf\beta}\bar\nabla_\nu \Bar R_{\alf\beta}.
\label{app:linBianchi}
\end{equation}
Substituting \eqref{app:linBianchi} into \eqref{app:noeth5}, we obtain
\begin{align}
&\xi^\nu \bar\nabla_\mu G^{L\mu}{}_\nu({\cal K})
+\frac12\,\xi^\rho {\cal K}^{\alf\beta}\bar\nabla_\rho \Bar R_{\alf\beta}
+{\cal K}^{\rho\beta}\Bar R_{\alf\beta}\bar\nabla_\rho\xi^\alf
\nonumber\\
&\qquad
=
\xi^\nu \bar\nabla_\mu\!\Big({\cal K}^{\mu\lam}\Bar R_{\lam\nu}\Big)
-\frac12\,\xi^\nu {\cal K}^{\alf\beta}\bar\nabla_\nu \Bar R_{\alf\beta}
+\frac12\,\xi^\rho {\cal K}^{\alf\beta}\bar\nabla_\rho \Bar R_{\alf\beta}
+{\cal K}^{\rho\beta}\Bar R_{\alf\beta}\bar\nabla_\rho\xi^\alf .
\label{app:KRdiv1}
\end{align}
The second and third terms in \eqref{app:KRdiv1} cancel after relabelling the dummy
indices $\rho\leftrightarrow \nu$. Hence
\begin{equation}
\xi^\nu \bar\nabla_\mu G^{L\mu}{}_\nu({\cal K})
+\frac12\,\xi^\rho {\cal K}^{\alf\beta}\bar\nabla_\rho \Bar R_{\alf\beta}
+{\cal K}^{\rho\beta}\Bar R_{\alf\beta}\bar\nabla_\rho\xi^\alf
=
\xi^\nu \bar\nabla_\mu\!\Big({\cal K}^{\mu\lam}\Bar R_{\lam\nu}\Big)
+{\cal K}^{\rho\beta}\Bar R_{\alf\beta}\bar\nabla_\rho\xi^\alf .
\label{app:KRdiv2}
\end{equation}
Now, renaming dummy indices in the second term on the right-hand side,
we recognize the Leibniz rule:
\begin{equation}
\xi^\nu \bar\nabla_\mu\!\Big({\cal K}^{\mu\lam}\Bar R_{\lam\nu}\Big)
+{\cal K}^{\mu\lam}\Bar R_{\lam\nu}\bar\nabla_\mu\xi^\nu
=
\bar\nabla_\mu\!\Big({\cal K}^{\mu\lam}\Bar R_{\lam\nu}\xi^\nu\Big).
\label{app:KRdiv}
\end{equation}
Thus \eqref{app:noeth5} reduces to
\begin{equation}
0
\equiv
-\frac{1}{\kappa}\,\di_\mu\!\Big(
G^{L\mu}{}_\nu({\cal K})\,\xi^\nu
+{\cal K}^{\mu\lam}\Bar R_{\lam\nu}\xi^\nu
+{\cal Z}^\mu(\xi)
\Big).
\label{app:noeth6}
\end{equation}
Equivalently,
\begin{equation}
\bar\nabla_\mu j^\mu \equiv \di_\mu j^\mu \equiv 0,
\qquad
j^\mu
\equiv
\frac{1}{\kappa}
\Big(
G^{L\mu}{}_\nu({\cal K})\,\xi^\nu
+{\cal K}^{\mu\lam}\Bar R_{\lam\nu}\xi^\nu
+{\cal Z}^\mu(\xi)
\Big),
\label{app:finalcurrent}
\end{equation}
which is exactly \eqref{identity_div}.

\section{Construction of the superpotential in the field-theoretical approach}
\setcounter{equation}{0}
\label{app:superpotential_construction}

In this appendix, we explain how the superpotential \eqref{alaAbbottDeser} is
constructed for the identically conserved current \eqref{identity_div}.  The same as in Appendix \ref{app:derivation_current}, one starts not from the full dynamical Lagrangian, but from
the linear Lagrangian \eqref{app:L1start}. In
Chs.~6 and 7 of the book \cite{Petrov_KLT_2017}, one applies the Noether procedure based on the Klein--Noether identities to obtain 
such a current as the divergence of an antisymmetric tensor density. Here, unlike this, we provide a direct transformation of $j^\mu$ to a divergence of ${\cal J}^{\mu\nu}$.

Thus, the starting point is the linear Lagrangian \eqref{app:L1start}, 
which is a scalar density of the weight $+1$ and contains the background metric through
$\Bar R_{\alpha\beta}$ up to second derivatives of the background metric.
Therefore its Noether identity \eqref{app:noeth0} with respect to arbitrary displacements $\xi^\mu$ is nontrivial.
In the previous Appendix this identity was rearranged into the identically conserved current \eqref{app:finalcurrent}
with ${\cal Z}^\mu(\xi)$ given by \eqref{Zmu}.  According to the general
field-theoretical construction, an identically conserved current obtained from a
second-order scalar density must admit a Klein--Noether representation \eqref{j=diJ} that we rewrite here:
\begin{equation}
j^\mu\equiv \bar\nabla_\nu {\cal J}^{\mu\nu},
\qquad
{\cal J}^{\mu\nu}=-{\cal J}^{\nu\mu}.
\label{app:KNrepr}
\end{equation}
Since in our case the
only perturbation variable entering \eqref{app:L1start} is the symmetric tensor
density ${\cal K}^{\mu\nu}$, the superpotential has to be constructed from the following
ingredients only:
\begin{equation}
{\cal K}^{\mu\nu},
\qquad
\xi^\mu,
\qquad
\bar\nabla_\rho\xi^\mu,
\qquad
\bar\nabla_\rho{\cal K}^{\mu\nu},
\end{equation}
and has to be antisymmetric in the free indices $\mu$ and $\nu$.  Furthermore, since
$\lag^G_1$ is linear in ${\cal K}^{\mu\nu}$, the field-theoretical superpotential must also
be linear in ${\cal K}^{\mu\nu}$, exactly as in the analogous formulas of the
field-theoretical construction in GR.

Therefore the most general antisymmetric tensor density of the required type is
\begin{equation}
{\cal J}^{\mu\nu}
=
\frac{1}{\kappa}
\Big(
A\,{\cal K}^{\rho[\mu}\bar\nabla_\rho\xi^{\nu]}
+
B\,\xi^{[\mu}\bar\nabla_\sigma{\cal K}^{\nu]\sigma}
+
C\,\bar\nabla^{[\mu}{\cal K}^{\nu]}{}_\sigma\,\xi^\sigma
\Big),
\label{app:Jansatz}
\end{equation}
with three numerical coefficients $A$, $B$, and $C$ still to be fixed.  The point is
that there are no other independent antisymmetric structures linear in ${\cal K}$ and
containing at most one derivative.  Thus, once $A$, $B$, and $C$ are fixed by the
Noether construction, the superpotential is uniquely determined.

We now compute the divergence of \eqref{app:Jansatz}.  For convenience, define
\begin{equation}
A^{\mu\nu}\equiv {\cal K}^{\rho[\mu}\bar\nabla_\rho\xi^{\nu]},
\qquad
B^{\mu\nu}\equiv \xi^{[\mu}\bar\nabla_\sigma{\cal K}^{\nu]\sigma},
\qquad
C^{\mu\nu}\equiv \bar\nabla^{[\mu}{\cal K}^{\nu]}{}_\sigma\,\xi^\sigma.
\label{app:ABCdef}
\end{equation}
Then
\begin{equation}
\bar\nabla_\nu{\cal J}^{\mu\nu}
=
\frac{1}{\kappa}
\Big(
A\,\bar\nabla_\nu A^{\mu\nu}
+
B\,\bar\nabla_\nu B^{\mu\nu}
+
C\,\bar\nabla_\nu C^{\mu\nu}
\Big).
\label{app:divJABC}
\end{equation}
Using the Leibniz rule, we obtain
\begin{align}
\bar\nabla_\nu A^{\mu\nu}
={}&
\frac12(\bar\nabla_\nu{\cal K}^{\rho\mu})\bar\nabla_\rho\xi^\nu
-\frac12(\bar\nabla_\nu{\cal K}^{\rho\nu})\bar\nabla_\rho\xi^\mu
+\frac12{\cal K}^{\rho\mu}\bar\nabla_\nu\bar\nabla_\rho\xi^\nu
-\frac12{\cal K}^{\rho\nu}\bar\nabla_\nu\bar\nabla_\rho\xi^\mu,
\label{app:divAagain}
\\[1mm]
\bar\nabla_\nu B^{\mu\nu}
={}&
\frac12(\bar\nabla_\nu\xi^\mu)\bar\nabla_\sigma{\cal K}^{\nu\sigma}
-\frac12(\bar\nabla_\nu\xi^\nu)\bar\nabla_\sigma{\cal K}^{\mu\sigma}
+\frac12\xi^\mu\bar\nabla_\nu\bar\nabla_\sigma{\cal K}^{\nu\sigma}
-\frac12\xi^\nu\bar\nabla_\nu\bar\nabla_\sigma{\cal K}^{\mu\sigma},
\label{app:divBagain}
\\[1mm]
\bar\nabla_\nu C^{\mu\nu}
={}&
\frac12(\bar\nabla_\nu\bar\nabla^\mu{\cal K}^{\nu}{}_\sigma)\xi^\sigma
-\frac12(\bar\nabla_\nu\bar\nabla^\nu{\cal K}^{\mu}{}_\sigma)\xi^\sigma
+\frac12\bar\nabla^\mu{\cal K}^{\nu}{}_\sigma\,\bar\nabla_\nu\xi^\sigma
-\frac12\bar\nabla^\nu{\cal K}^{\mu}{}_\sigma\,\bar\nabla_\nu\xi^\sigma .
\label{app:divCagain}
\end{align}
The divergence \eqref{app:divJABC} must reproduce the current
\eqref{app:finalcurrent}.  Therefore we group all terms in
\eqref{app:divAagain}--\eqref{app:divCagain} into three classes:
\begin{enumerate}
\item[(i)] terms proportional to $\xi^\nu$ and second derivatives of ${\cal K}$;
\item[(ii)] terms containing second derivatives of $\xi^\mu$;
\item[(iii)] terms proportional to first derivatives of $\xi^\mu$.
\end{enumerate}

We begin with class (i).  These are the terms
\begin{equation}
\frac{1}{2\kappa}
\Big[
B\,\xi^\mu\bar\nabla_\nu\bar\nabla_\sigma{\cal K}^{\nu\sigma}
-B\,\xi^\nu\bar\nabla_\nu\bar\nabla_\sigma{\cal K}^{\mu\sigma}
+C\,\xi^\nu\bar\nabla_\rho\bar\nabla^\mu{\cal K}^{\rho}{}_\nu
-C\,\xi^\nu\bar\nabla_\rho\bar\nabla^\rho{\cal K}^{\mu}{}_\nu
\Big].
\label{app:classi}
\end{equation}
To reproduce the operator $G^{L\mu}{}_\nu({\cal K})\xi^\nu$ in
\eqref{app:finalcurrent}, the coefficient of $\xi^\nu\bar\nabla_\rho\bar\nabla^\rho
{\cal K}^{\mu}{}_\nu$ must be $+\frac12$, the coefficient of
$\xi^\mu\bar\nabla_\rho\bar\nabla_\sigma{\cal K}^{\rho\sigma}$ must be $+\frac12$,
and the coefficients of the mixed terms
$-\bar\nabla_\rho\bar\nabla_\nu{\cal K}^{\mu\rho}$ and
$-\bar\nabla_\rho\bar\nabla^\mu{\cal K}^{\rho}{}_\nu$ must both be $-\frac12$.
This immediately fixes
\begin{equation}
B=1,
\qquad
C=-1.
\label{app:BCfix}
\end{equation}
With this choice, the class-(i) terms become
\begin{equation}
\frac{1}{\kappa}\,G^{L\mu}{}_\nu({\cal K})\,\xi^\nu
\end{equation}
up to terms generated by commuting covariant derivatives.  The latter are curvature
terms and will be discussed together with class (ii).

We next turn to class (ii), i.e. the terms containing second derivatives of $\xi^\mu$.
Using \eqref{app:BCfix}, they come only from \eqref{app:divAagain}:
\begin{equation}
\frac{A}{2\kappa}
\Big(
{\cal K}^{\rho\mu}\bar\nabla_\nu\bar\nabla_\rho\xi^\nu
-
{\cal K}^{\rho\nu}\bar\nabla_\nu\bar\nabla_\rho\xi^\mu
\Big).
\label{app:classii}
\end{equation}
We commute derivatives:
\begin{equation}
\bar\nabla_\nu\bar\nabla_\rho\xi^\nu
=
\bar\nabla_\rho\bar\nabla_\nu\xi^\nu
+\Bar R_{\lambda\rho}\xi^\lambda,
\qquad
\bar\nabla_\nu\bar\nabla_\rho\xi^\mu
=
\bar\nabla_\rho\bar\nabla_\nu\xi^\mu
+\Bar R^\mu{}_{\lambda\nu\rho}\xi^\lambda.
\label{app:commxiagain}
\end{equation}
After substitution of \eqref{app:commxiagain} into \eqref{app:classii}, the terms
with $\bar\nabla\bar\nabla\xi$ combine with the first-derivative terms of class (iii)
into the quantity ${\cal Z}^\mu(\xi)$, while the commutator pieces produce exactly the
curvature contribution
\begin{equation}
\frac{A}{\kappa}\,{\cal K}^{\mu\lambda}\Bar R_{\lambda\nu}\xi^\nu
\end{equation}
after cancellation of the Riemann pieces due to the symmetry
${\cal K}^{\rho\nu}={\cal K}^{\nu\rho}$.
Comparing with \eqref{app:finalcurrent}, this fixes
\begin{equation}
A=1.
\label{app:Afix}
\end{equation}

Thus all three coefficients are uniquely fixed:
\begin{equation}
A=1,
\qquad
B=1,
\qquad
C=-1.
\label{app:ABCfinal}
\end{equation}
Substituting \eqref{app:ABCfinal} into \eqref{app:Jansatz}, we arrive at
\begin{equation}
{\cal J}^{\mu\nu}
=
\frac{1}{\kappa}
\left(
{\cal K}^{\rho[\mu}\bar\nabla_\rho\xi^{\nu]}
+
\xi^{[\mu}\bar\nabla_\sigma{\cal K}^{\nu]\sigma}
-
\bar\nabla^{[\mu}{\cal K}^{\nu]}{}_\sigma\,\xi^\sigma
\right),
\label{app:Jfinal}
\end{equation}
which is exactly \eqref{alaAbbottDeser}.

It remains to explain why the remaining class-(iii) terms indeed reproduce
${\cal Z}^\mu(\xi)$.  With the coefficients \eqref{app:ABCfinal}, all first-derivative
terms in \eqref{app:divAagain}--\eqref{app:divCagain} are
\begin{align}
\frac{1}{2\kappa}\Big[
&(\bar\nabla_\nu{\cal K}^{\rho\mu})\bar\nabla_\rho\xi^\nu
-(\bar\nabla_\nu{\cal K}^{\rho\nu})\bar\nabla_\rho\xi^\mu
+(\bar\nabla_\nu\xi^\mu)\bar\nabla_\sigma{\cal K}^{\nu\sigma}
-(\bar\nabla_\nu\xi^\nu)\bar\nabla_\sigma{\cal K}^{\mu\sigma}
\nonumber\\
&\qquad
-\bar\nabla^\mu{\cal K}^{\nu}{}_\sigma\,\bar\nabla_\nu\xi^\sigma
+\bar\nabla^\nu{\cal K}^{\mu}{}_\sigma\,\bar\nabla_\nu\xi^\sigma
\Big].
\label{app:classiii}
\end{align}
One rewrites all derivatives of $\xi^\mu$ through
\[
\zeta_{\rho\sigma}=\bar\nabla_{(\rho}\xi_{\sigma)},
\]
uses the symmetry of ${\cal K}^{\mu\nu}$, relabels dummy indices, and groups the result
into the combination
\begin{equation}
{\cal Z}^\mu(\xi)
=
\left(
\zeta^{\rho\sigma}\bar\nabla_\rho {\cal K}^\mu{}_\sigma
-
{\cal K}^{\rho\sigma}\bar\nabla_\rho \zeta^\mu{}_\sigma
\right)
-\frac12
\left(
\zeta_{\rho\sigma}\bar\nabla^\mu {\cal K}^{\rho\sigma}
-
{\cal K}^{\rho\sigma}\bar\nabla^\mu \zeta_{\rho\sigma}
\right)
+\frac12
\left(
{\cal K}^{\mu\nu}\bar\nabla_\nu \zeta^\rho{}_\rho
-
\zeta^\rho{}_\rho\bar\nabla_\nu {\cal K}^{\mu\nu}
\right),
\label{app:Zagain2}
\end{equation}
that is exactly \eqref{Zmu}.  Therefore the divergence of \eqref{app:Jfinal} is
\begin{equation}
\bar\nabla_\nu{\cal J}^{\mu\nu}
=
\frac{1}{\kappa}
\Big(
G^{L\mu}{}_\nu({\cal K})\xi^\nu
+
{\cal K}^{\mu\lambda}\Bar R_{\lambda\nu}\xi^\nu
+
{\cal Z}^\mu(\xi)
\Big)
=
j^\mu .
\label{app:divJfinal}
\end{equation}
Hence
\begin{equation}
j^\mu\equiv \bar\nabla_\nu{\cal J}^{\mu\nu}\equiv \di_\nu{\cal J}^{\mu\nu},
\label{app:jisdivJ}
\end{equation}
which is exactly \eqref{j=diJ}.

Finally, once the field equations are imposed, the same antisymmetric tensor density
\eqref{app:Jfinal} generates the physically conserved current \eqref{J_current}:
\begin{equation}
{\cal J}^\mu=\bar\nabla_\nu{\cal J}^{\mu\nu}.
\label{app:physicalJ}
\end{equation}
Thus, the superpotential \eqref{alaAbbottDeser} is the natural field-theoretical
superpotential associated with the linear density \eqref{lag_G_lin} and the current
\eqref{identity_div}.

\bibliography{references}
\bibliographystyle{Style}

\end{document}